\documentclass[11pt,letterpaper]{article}
\pdfoutput=1
\usepackage{soul}
\usepackage{jheppub}
\usepackage{amsfonts}
\usepackage{amsmath}
\allowdisplaybreaks[4]         
\usepackage{amssymb}   
\usepackage{euscript}       
\usepackage{color}
\usepackage[normalem]{ulem}
\usepackage{mathrsfs}  
\usepackage{booktabs}
\usepackage{siunitx}
\usepackage{amsmath,latexsym,xcolor,mathtools}
\usepackage{graphicx,verbatim, subcaption, caption,inputenc}

\newcommand{\bea}{\begin{eqnarray}}
\newcommand{\eea}{\end{eqnarray}}
\newcommand{\ba}{\begin{eqnarray}}
\usepackage{braket}
\newcommand{\ea}{\end{eqnarray}}

\newcommand{\beq}{\begin{equation}}
\newcommand{\eeq}{\end{equation} }
\newcommand{\beqa}{\begin{eqnarray}}

\newcommand{\eeqa}{\end{eqnarray}}
\newcommand{\beqar}{\begin{eqnarray*}}
\newcommand{\eeqar}{\end{eqnarray*}}

\newcommand{\be}{\begin{equation}}
\newcommand{\ee}{\end{equation}}

\begin{document}

\title{Renormalized Holographic Entanglement Entropy in Lovelock Gravity}
\author[a]{Giorgos Anastasiou,}
\author[b]{Ignacio J. Araya,} 
\author[c]{Robert B. Mann}
\author[d]{and Rodrigo Olea}

\emailAdd{georgios.anastasiou@pucv.cl}
\emailAdd{ignaraya@unap.cl}
\emailAdd{rbmann@uwaterloo.ca}
\emailAdd{rodrigo.olea@unab.cl}
\affiliation[a]{Instituto de F\'{\i}sica, Pontificia Universidad Cat\'{o}lica de Valpara\'{\i}so, Casilla 4059, Valpara\'{\i}so, Chile}
\affiliation[b]{Instituto de Ciencias Exactas y Naturales, Facultad de Ciencias, Universidad Arturo Prat, Avenida Arturo Prat Chac\'on 2120, 1110939, Iquique, Chile}
\affiliation[c]{Department of Physics and Astronomy, University of Waterloo, Waterloo, Ontario, N2L 3G1, Canada}
\affiliation[d]{Departamento de Ciencias F\'{\i}sicas, Universidad Andres Bello, Sazi\'{e} 2212, Piso 7, Santiago, Chile}

\abstract{
We study the renormalization of Entanglement Entropy in holographic CFTs dual
to Lovelock gravity. It is known that the holographic EE in Lovelock gravity
is given by the Jacobson-Myers (JM) functional. As usual, due to the divergent
Weyl factor in the Fefferman-Graham expansion of the boundary metric for
Asymptotically AdS spaces, this entropy functional is infinite. By considering
the Kounterterm renormalization procedure, which utilizes extrinsic boundary
counterterms in order to renormalize the on-shell Lovelock gravity action for
AAdS spacetimes, we propose a new renormalization prescription for the
Jacobson-Myers functional. We then explicitly show the cancellation of
divergences in the EE up to next-to-leading order in the holographic radial
coordinate, for the case of spherical entangling surfaces. Using this new
renormalization prescription, we directly find the $C-$function candidates for
odd and even dimensional CFTs dual to Lovelock gravity. Our results illustrate the  notable improvement that the Kounterterm method  affords  over other approaches, as it is non-perturbative and does not require that the Lovelock theory has limiting Einstein behavior.}
 
\maketitle

\section{Introduction}

Entanglement entropy has played an significant role in advancing our understanding of holography.  The 
Ryu-Takayanagi (RT) prescription \cite{Ryu:2006bv} allows one to compute the entanglement entropy for a region $A$ in a conformal field
theory (CFT) dual to Einstein gravity (and additional matter fields) in terms of the area $\mathcal{A}(\Sigma_A)$ of a surface $\Sigma_A$ of minimal area in the bulk  that is homologous to $A$
\begin{equation}\label{SEE}
S_{\textrm{EE}}(A) = \frac{\mathcal{A}(\Sigma_A) }{4G}
\end{equation}
known as the Holographic Entanglement Entropy (HEE) formula,
where $G$ is the gravitational constant.     

Higher curvature terms likewise enrich our understanding of holography. They  appear as  quantum (or stringy) corrections to Einstein gravity \cite{Grisaru:1986px,Gross:1986iv} and in general have holographic duals that are not equivalent to those defind from Einstein gravity \cite{Gubser:1998bc}.  They have been used to investigate interesting CFT physics 
 \cite{Buchel:2009sk,deBoer:2009gx,Myers:2010jv,Camanho:2013pda,Sinamuli:2017rhp,Bueno:2018xqc}
 that in some cases is quite universal, applicable to very general CFTs \cite{Kats:2007mq,Brigante:2007nu,Camanho:2009vw,Camanho:2010ru,Myers:2010tj,Mezei:2014zla,Bueno:2015rda,Miao:2015dua,Bueno:2018yzo,Bueno:2020odt}. 

Higher-curvature terms generalize both  boundary terms of the action functional \cite{Teitelboim:1987zz,Myers:1987yn,Davis:2002gn,Grumiller:2008ie}  and the  Bekenstein- Hawking black hole entropy-area relation \cite{Bekenstein:1973ur,Hawking:1974sw} in terms of the Wald formula \cite{Wald:1993nt,Iyer:1994ys}, and it is natural to expect they will modify the entanglement entropy formula in \eqref{SEE}.  However employing the expected Wald functional fails \cite{Hung:2011xb} because  the extrinsic curvature(s) of the generalized bulk surface must be taken into account.  This was first carried out for quadratic gravity  \cite{Fursaev:2013fta}, and then a general formula was obtained for theories whose actions have  arbitrary contractions of the Riemann tensor \cite{Dong:2013qoa,Camps:2013zua,Bhattacharyya:2013gra,Bhattacharyya:2013jma}.  
However this formula involved taking a weighted sum over trace anomaly charges whose evaluation, beyond quadratic order, entails a  theory-dependent splitting of the Riemann tensor components that is somewhat complicated.  This shortcoming was recently circumvented \cite{Bueno:2020uxs} in terms of a general formula obtained  in terms of implicit derivatives of a Euclidean higher curvature action with respect to  projections of the Riemann tensor and extrinsic curvature tensors associated with the RT surface $\Sigma_A$.

We obtain in this paper the renormalization  of the HEE formula, suitable for any Lovelock theory of gravity \cite{Lovelock:1971yv}, which is
\begin{equation}
S_{EE}^{ren}(A) =S_{JM}\left[  \Sigma_A\right]  +\frac{c_{d}\left\lfloor \frac
{d+1}{2}\right\rfloor }{4G}%
{\displaystyle\int\limits_{\partial\Sigma}}
B_{d-2}, \label{S_ren_intro}%
\end{equation}
where $\Sigma_A$ is the codimension-2 surface that extremizes the Jacobson-Myers
(JM) functional $S_{JM}\left[  \Sigma\right] $  \cite{Jacobson:1993xs}, $c_{d}$ is a
dimension-dependent constant  and
$B_{d-2}$ is an extrinsic boundary counterterm (both given below in eqs.\eqref{KountertermCoupling} and \eqref{Kounterterm}) and $\left\lfloor x\right\rfloor $ is the usual floor function. We obtain this by making use of the Kounterterm renormalization procedure \cite{Olea:2005gb,Olea:2006vd,Kofinas:2007ns}.   This procedure has been   successfully
applied to   computations for holographic CFTs dual to Einstein-AdS
gravity \cite{Anastasiou:2017xjr,Anastasiou:2018rla,Anastasiou:2018mfk,Anastasiou:2019ldc}. 
It was recently used  to obtain an expression for the conserved charges of solutions having $k$-fold degenerate vacua in Lovelock AdS gravity, making manifest a link between the degeneracy of a given vacuum and the nonlinearity of the energy formula \cite{Arenas-Henriquez:2019rph,Arenas-Henriquez:2017xnr}.

The prescription \eqref{S_ren_intro} exploits  the replica formula, whose holographic implementation
\cite{Lewkowycz:2013nqa,Nishioka:2018khk} generalized the   minimal area prescription \eqref{SEE} beyond the spherically-symmetric case \cite{Casini:2011kv}.  The entanglement entropy in the saddle-point
approximation of AdS/CFT is given by 
\begin{equation}
S_{EE}=-\lim_{\alpha\rightarrow1}\partial_{\alpha}I_{E}\left[  M_{d+1}%
^{\left(  \alpha\right)  }\right]  ,
\label{Replica_EE}
\end{equation}
where $I_{E}\left[  M_{d+1}^{\left(  \alpha\right)  }\right]  $ is the
Euclidean on-shell action for the bulk, evaluated on a suitably constructed
$\left(  d+1\right)  -$dimensional conically singular orbifold $M_{d+1}%
^{\left(  \alpha\right)  }$ (with angular deficit given by $2\pi\left(
1-\alpha\right)  $).   In the case of
Einstein-AdS \cite{Dong:2013qoa}, $M_{d+1}^{\left(  \alpha\right)  }$ is the backreacted manifold
sourced by a codimension-2 cosmic brane with tension $T=\frac{\left(
1-\alpha\right)  }{4G}$, coupled to the ambient geometry through the
Nambu-Goto (NG) action. In the $\alpha\rightarrow1$ tensionless limit, the NG
action decouples from the ambient geometry and the extremal codimension-2
surface that represents the on-shell location of the brane becomes the usual
RT surface.

The computation of HEE is therefore directly related to the
evaluation of on-shell gravity actions, such that the divergences in the
former are entirely due to the divergences in the latter. It then becomes
evident that if one considers the renormalized on-shell action, the entropies
thus computed will be renormalized as well.

In the Kounterterm approach,  renormalization of the on-shell Einstein-AdS action is carried
out by considering extrinsic boundary counterterms (Kounterterms) \cite{Olea:2005gb,Olea:2006vd}. The
usual asymptotic charges and thermodynamic behavior of asymptotically AdS (AAdS) black-hole
solutions are correctly recovered. Furthermore, the agreement of this extrinsic
counterterm renormalization procedure with the standard Holographic
Renormalization scheme \cite{deHaro:2000vlm} has been demonstrated
for a large class of  AAdS spaces \cite{Anastasiou:2020zwc}.  The Kounterterm-renormalized action has been evaluated on  $M_{d+1}^{\left(  \alpha\right)  }$ orbifolds \cite{Anastasiou:2018mfk}, based on the work of Fursaev, Patrushev and Solodukhin \cite{Fursaev:2013fta}. In doing so, via  eq.\eqref{Replica_EE}, the renormalized HEE was readily computed.

We are interested here in   holographic CFTs that are dual to Lovelock gravity.  In
order to compute the holographic EE in this case, we note that eq.\eqref{Replica_EE} is independent of the particular type of dual gravity
theory. This is because \eqref{Replica_EE}  assumes only that the AdS/CFT
correspondence holds, namely that the gravitational theory of choice is the correct
dual of the CFT under study. This, in turn, implies that the saddle-point approximation is
valid, such that the partition function of the CFT is given by the
exponential of (minus) the Euclidean on-shell gravity action of the dual bulk
manifold. Then, one need only evaluate $I_{E}\left[  M_{d+1}^{\left(
\alpha\right)  }\right]  $  for the corresponding gravity theory and
\eqref{Replica_EE} still applies.
We thereby obtain eq.\eqref{S_ren_intro} using logic similar to that in Einstein-AdS gravity,
employing the   renormalization of the on-shell Lovelock gravity action developed  in refs.\cite{Kofinas:2007ns,Kofinas:2008ub}, and applying it to  
conically singular manifolds \cite{Anastasiou:2018mfk}.

We consider even and odd dimensional CFTs separately.
For   odd dimensional CFTs, the renormalized HEE in eq.\eqref{S_ren_intro} can be rewritten in terms of the intrinsic AdS
curvature $\mathcal{F}$ of the minimal surface $\Sigma_A$ 
(defined below in eq.\eqref{F_AdS_cod2}) and the Euler characteristic of $\Sigma_A$. For
spherical entangling surfaces only the topological number contributes (as we will show in section \ref{BallHEE}). In both odd and even dimensional CFTs, the HEE
counterterm given by the $B_{d-2}$ term in eq.\eqref{S_ren_intro} is
explicitly shown to cancel the leading order divergence coming from the JM
functional. It also cancels the next-to-leading order divergence in the case of
spherical entangling surfaces in conformally flat AdS boundaries. The
renormalized EE obtained here corresponds to the finite part for
odd-dimensional CFTs and the logarithmically divergent part for  even-dimensional CFTs. In both cases these quantities are universal, as they are related to the
holographic $C-$function candidate of the CFT, which for the odd-$d$
case is the $a^{\ast}$ charge \cite{Myers:2010xs} and for the even-$d$ case is
the type-$A$ anomaly coefficient  \cite{Dong:2013qoa,Camps:2013zua}.

Our paper is organized as follows. In section \ref{Preliminaries}, we
review the Lovelock-AdS gravity theory, emphasizing the equation of motion (EOM) and
its factorization in terms of the different maximally-symmetric configurations
for the vacua of the theory. In section \ref{RenormalizedLovelock} we revisit
the renormalized Lovelock-AdS action \cite{Kofinas:2007ns}, and in the
case of even-dimensional bulk manifolds, we rewrite the action in terms of a
polynomial $P\left(\mathcal{F}\right)  $ in the AdS curvature with respect
to a chosen vacuum \cite{Kastor:2006vw}. In section \ref{ActionOnOrbifold}, we
evaluate the renormalized Lovelock-AdS action on the replica orbifold, in
order to obtain the contribution of the co-dimension 2 extremal surface to the action,
and from there, the renormalized HEE. We also review the derivation of the
Euler-Lagrange equation for the extremal surface of the JM functional, which
defines the co-dimension 2 surface whose JM entropy gives the HEE. We show how to write
this equation in a new factorized form. In section
\ref{DivergenceCancellation}, we exhibit the explicit cancellation of leading
and next-to-leading order divergences in the renormalized HEE. In section
\ref{TopologicalHEE}, we consider odd-dimensional holographic CFTs dual to
Lovelock theory and we decompose the HEE into a geometric part written as a
polynomial on the AdS curvature $\mathcal{F}$ of the intrinsic metric of the
extremal surface $\Sigma$, and a purely topological part that depends on the
Euler characteristic of $\Sigma$. In section \ref{BallHEE}, we consider the
example of ball-shaped entangling regions in the CFT and we compute the
renormalized HEE for both odd-dimensional and even-dimensional CFTs, relating
the resulting universal part to the $a^{\ast}$ charge (or generalized $F$
quantity) and type-A anomaly coefficient respectively, both of which are
$C-$function candidates. In section \ref{Conclusions}, we summarize our results
for the HEE and relate them to the holographic properties of the CFT. We also
give some general conclusions based on our results and discuss possible future
avenues of research.

\section{Preliminaries: Lovelock Gravity and factorized equations of motion}
\label{Preliminaries}

Lovelock gravity is the most general pure gravity action such that it has
second order differential equations for the dynamical variable, i.e., the metric \cite{Lovelock:1971yv}. The
Lovelock action is given by%
\begin{equation}
I_{L}\left[  M_{d+1}\right]  =\frac{1}{16\pi G}%
{\displaystyle\int\limits_{M_{d+1}}}
d^{d+1}x%
{\displaystyle\sum\limits_{p=0}^{\left\lfloor \frac{d}{2}\right\rfloor }}
\alpha_{p}L_{2p}, \label{Lovelock_Action}%
\end{equation}
where the Lovelock densities $L_{2p}$ are defined by%
\begin{equation}
L_{2p}=\frac{1}{2^{p}}\sqrt{-\mathcal{G}}\delta_{\mu_{1}\cdots\mu_{2p}}%
^{\nu_{1}\cdots\nu_{2p}}R_{\phantom{\mu_{1}\mu_{2}}\nu_{1}\nu_{2}}^{\mu_{1}\mu_{2}}\cdots
R_{\phantom{\mu_{2p-1}\mu_{2p}}\nu_{2p-1}\nu_{2p}}^{\mu_{2p-1}\mu_{2p}}, \label{Lovelock_Density}%
\end{equation}
$\delta_{\mu_{1}\cdots\mu_{2p}}%
^{\nu_{1}\cdots\nu_{2p}}$ is the generalized Kronecker-delta, 
and $\left\lfloor x\right\rfloor $ is the integer floor of $x$. We note that
Einstein-AdS gravity is a particular case of the Lovelock action defined in
eq.\eqref{Lovelock_Action}, for which%
\begin{align}
\alpha_{0}  &  =-2\Lambda=\frac{d\left(  d-1\right)  }{\ell^{2}}\,, \qquad \alpha_{1}    =1\,,
\label{Einstein-AdS}%
\end{align}
and $\alpha_{i}=0$ for $i>1$. The Lovelock theories we shall consider  are higher-curvature corrections to Einstein-AdS gravity and therefore, unless otherwise stated, the values of $\alpha_{0}$ and $\alpha_{1}$ are
always given as in eq.\eqref{Einstein-AdS}.

The Lovelock EOM is given by \cite{Lovelock:1971yv,Arenas-Henriquez:2017xnr}

\begin{equation}
E_{\mu}^{\nu}=%
{\displaystyle\sum\limits_{p=0}^{\left\lfloor \frac{d}{2}\right\rfloor }}
\frac{\alpha_{p}}{2^{p+1}}\delta_{\mu\mu_{1}\cdots\mu_{2p}}^{\nu\nu_{1}%
\cdots\nu_{2p}} R_{\phantom{\mu_{1}\mu_{2}}\nu_{1}\nu_{2}}^{\mu_{1}\mu_{2}}\cdots
R_{\phantom{\mu_{2p-1}\mu_{2p}}\nu_{2p-1}\nu_{2p}}^{\mu_{2p-1}\mu_{2p}} =0\,, \label{LovelockEOM}%
\end{equation}
and using the values of $\alpha_{0}$ and $\alpha_{1}$ given in
eq.\eqref{Einstein-AdS}, it can be rewritten as

\begin{gather}
R_{\nu}^{\mu}-\frac{1}{2}\left(  R-2\Lambda\right)  \delta_{\nu}^{\mu}=H_{\nu
}^{\mu},\\
H_{\nu}^{\mu}=%
{\displaystyle\sum\limits_{p=2}^{\left\lfloor \frac{d}{2}\right\rfloor }}
\frac{\alpha_{p}}{2^{p+1}}\delta_{\nu\mu_{1}\cdots\mu_{2p}}^{\mu\nu_{1}%
\cdots\nu_{2p}} R_{\phantom{\mu_{1}\mu_{2}}\nu_{1}\nu_{2}}^{\mu_{1}\mu_{2}}\cdots
R_{\phantom{\mu_{2p-1}\mu_{2p}}\nu_{2p-1}\nu_{2p}}^{\mu_{2p-1}\mu_{2p}}\,,
\end{gather}
where $H_{\nu}^{\mu}$ is the Lanczos-Lovelock tensor.

As usual, by considering the maximally-symmetric (constant
curvature) ansatz for the Riemann curvature tensor in AdS, given by%
\begin{equation}
R_{\phantom{\mu_{1}\mu_{2}}\nu_{1}\nu_{2}}^{\mu_{1}\mu_{2}} =-\frac{1}{\ell_{\text{eff}}^{2}}%
\delta_{\nu_{1}\nu_{2}}^{\mu_{1}\mu_{2}}\,,
\end{equation}
and inserting it into the EOM  \eqref{LovelockEOM}, we
obtain a condition for $\ell_{\text{eff}}^{-2}$ 
\begin{equation}
\Delta\left(  \ell_{\text{eff}}^{-2}\right)  \overset{\text{def.}}{=}%
{\displaystyle\sum\limits_{p=0}^{\left\lfloor \frac{d}{2}\right\rfloor }}
\frac{\left(  -1\right)  ^{p+1}\left(  d-2\right)  !\alpha_{p}}{\left(
d-2p\right)  !}\left(  \frac{1}{\ell_{\text{eff}}^{2}}\right)  ^{p}=0 \,,
\label{CharPoly}%
\end{equation}
given in terms of the
characteristic polynomial of the theory \cite{Arenas-Henriquez:2017xnr}.
Thus, the roots of   $\Delta\left(  \ell_{\text{eff}}^{-2}\right)  =0$
give the possible effective AdS radii for the vacua of the Lovelock theory
characterized by the set of $\left\{  \alpha_{p}\right\}  $ couplings.

It is easy to see that the roots obtained from eq.\eqref{CharPoly}  may have algebraic multiplicity higher than one,
which in turn implies that the vacua of the corresponding Lovelock theory are
degenerate. By simple algebra considerations, the $k-$th degeneracy condition
is defined as
\begin{equation}
\Delta^{\left(  k\right)  }=\frac{1}{k!}\frac{d^{k}\Delta}{d\left(
\ell_{\text{eff}}^{-2}\right)  ^{k}}=%
{\displaystyle\sum\limits_{p=k}^{\left\lfloor \frac{d}{2}\right\rfloor }}
\frac{\left(  -1\right)  ^{p+1}\left(  d-2\right)  !p!\alpha_{p}}{k!\left(
p-k\right)  !\left(  d-2p\right)  !}\left(  \frac{1}{\ell_{\text{eff}}^{2}%
}\right)  ^{p-k}=0\,, \label{k-thDeg}%
\end{equation}
in agreement with ref.\cite{Arenas-Henriquez:2017xnr}. A theory is $\left(  k-1\right)
-$degenerate if the largest algebraic multiplicity of its vacua is $k$, which
in turn means that all $\Delta^{\left(  q\right)  }$ for $q<k$ are zero. 
Note that the normalization of the degeneracy conditions as considered in
eq.\eqref{k-thDeg} is such that the value of $\Delta^{\left(  1\right)  }$ for
Einstein-AdS gravity is equal to one.

As shown in Appendix \ref{AppendixB}, the EOM \eqref{LovelockEOM} can be
rewritten in factorized form as
\begin{align}
E_{\nu}^{\mu}&=\frac{\alpha_{N}}{2^{N+1}}\delta_{\nu\mu_{1}\cdots\mu_{2N}}%
^{\mu\nu_{1}\cdots\nu_{2N}}\left(  R_{\phantom{\mu_{1}\mu_{2}}\nu_{1}\nu_{2}}^{\mu_{1}\mu_{2}}%
+ \frac{1}{\ell_{\text{eff}(1)}^{2}}\delta_{\nu_{1}\nu_{2}}^{\mu_{1}\mu_{2}%
}\right)  \cdots\left(  R_{\phantom{\mu_{2N-1}\mu_{2N}}\nu_{2N-1}\nu_{2N}}^{\mu_{2N-1}\mu_{2N}}+ \frac
{1}{\ell_{\text{eff}(N)}^{2}}\delta_{\nu_{2N-1}\nu_{2N}}^{\mu_{2N-1}\mu_{2N}%
}\right)\nonumber\\
& =0 \,, 
\label{LovelockEomFactorization}%
\end{align}
where $N\leq\left\lfloor \frac{d}{2}\right\rfloor $ is the order (in powers of
the Riemann curvature) of the Lovelock Lagrangian and $\left\{  \ell
_{\text{eff}(i)}\right\}  $ are the effective AdS radii of the theory, given
by the solutions of the characteristic polynomial of eq.\eqref{CharPoly} \cite{Kastor:2006vw}. When
the theory has $\left(  k-1\right)  -$degenerate vacua, the term corresponding
to the $i-$th degenerate vacuum is repeated $k$ times in the product.

Having reviewed the equations of motion for Lovelock gravity theories, we proceed in the next section with their renormalization. Especially in the case of
even-dimensional bulks, a useful rewriting of the renormalized action in terms
of a polynomial on the AdS curvature of the manifold is obtained, which is a
generalization of the renormalized volume formula proposed for Einstein-AdS 
\cite{Anastasiou:2018mfk}. 
 
\section{Renormalized Lovelock-AdS action}
\label{RenormalizedLovelock}

We consider the renormalized Lovelock-AdS action, given by \cite{Kofinas:2007ns}
\begin{equation}
I_{L}^{ren}\left[  M_{d+1}\right]  =\frac{1}{16\pi G}%
{\displaystyle\int\limits_{M_{d+1}}}
{\displaystyle\sum\limits_{p=0}^{\left\lfloor \frac{d}{2}\right\rfloor }}
\alpha_{p}L_{2p}+\frac{c_{d}}{16\pi G}%
{\displaystyle\int\limits_{\partial M_{d+1}}}
d^{d}x B_{d}, \label{KountertermLovelockAction}%
\end{equation}
where the $L_{2p}$ Lovelock densities are defined in
eq.\eqref{Lovelock_Density}, and the coupling $c_{d}$ is defined by%
\begin{equation}
c_{d}=\left\{
\begin{tabular}
[c]{l}%
$\frac{2}{d+1}%
{\displaystyle\sum\limits_{p=1}^{\frac{d-1}{2}}}
\frac{p\alpha_{p}\left(  -1\right)  ^{\frac{d+3}{2}-p}}{\left(  d+1-2p\right)
!}\left(  \ell_{\text{eff}}\right)  ^{\left(  d+1-2p\right)  }$ \qquad\qquad\qquad\quad for odd $d$\\
$\frac{2}{d}\left[  \frac{\left(  d-1\right)  !}{2^{d-2}\left[  \left(
\frac{d}{2}-1\right)  !\right]  ^{2}}\right]
{\displaystyle\sum\limits_{p=1}^{\frac{d}{2}}}
\frac{p\alpha_{p}\left(  -1\right)  ^{\frac{d}{2}+1-p}}{\left(  d+1-2p\right)
!}\left(  \ell_{\text{eff}}\right)  ^{\left(  d-2p\right)  }$ \qquad for even $d$%
\end{tabular}
\ \ \ \ \ \right.  . \label{KountertermCoupling}%
\end{equation}
Here, $\ell_{\text{eff}}$ is the effective AdS radius of the branch under consideration. Also, the boundary Kounterterm  is given by 
\begin{equation}
B_{d}=\left\{
\begin{array}
[c]{c}%
\underset{\text{ for odd }d}{-\left(  d+1\right)  \sqrt{-h}%
{\displaystyle\int\limits_{0}^{1}}
dt\delta_{i_{1}\cdots i_{d}}^{j_{1}\cdots j_{d}}K_{j_{1}}^{i_{1}}\left(
\frac{1}{2}\mathcal{R}_{\phantom{i_{2}i_{3}}j_{2}j_{3}}^{i_{2}i_{3}}-t^{2}K_{j_{2}}^{i_{2}%
}K_{j_{3}}^{i_{3}}\right)  \cdots\left(  \frac{1}{2}\mathcal{R}_{\phantom{i_{d-1}i_{d}}j_{d-1}j_{d}%
}^{i_{d-1}i_{d}}-t^{2}K_{j_{d-1}}^{i_{d-1}}K_{j_{d}}^{i_{d}}\right)  }\\
-d\sqrt{-h}%
{\displaystyle\int\limits_{0}^{1}}
dt%
{\displaystyle\int\limits_{0}^{t}}
ds\delta_{i_{1}\cdots i_{d-1}}^{j_{1}\cdots j_{d-1}}K_{j_{1}}^{i_{1}}\left(
\frac{1}{2} \mathcal{R}_{\phantom{i_{2}i_{3}}j_{2}j_{3}}^{i_{2}i_{3}}-t^{2}K_{j_{2}}^{i_{2}%
}K_{j_{3}}^{i_{3}}+\frac{s^{2}}{\ell_{\text{eff}}^{2}}\delta_{j_{2}}^{i_{2}%
}\delta_{j_{3}}^{i_{3}}\right)  \times\\
\underset{\text{for even }d}{\qquad\cdots\left(  \frac{1}{2}\mathcal{R}%
_{\phantom{i_{d-2}i_{d-1}}j_{d-2}j_{d-1}}^{i_{d-2}i_{d-1}}-t^{2}K_{j_{d-2}}^{i_{d-2}}K_{j_{d-1}%
}^{i_{d-1}}+\frac{s^{2}}{\ell_{\text{eff}}^{2}}\delta_{j_{d-2}}^{i_{d-2}%
}\delta_{j_{d-1}}^{i_{d-1}}\right)  }%
\end{array}
\right.  , \label{Kounterterm}%
\end{equation}
where $K_{j}^{i}$ is the extrinsic curvature of the foliation with respect to
the radial coordinate $\rho$ and $\mathcal{R}_{\phantom{i_{1}i_{2}}j_{1}j_{2}}%
^{i_{1}i_{2}}$ is the Riemann curvature of the intrinsic metric $h$ in the
foliation \cite{Kofinas:2007ns,Kofinas:2008ub}. We emphasize that $\ell_{\text{eff}}$ in
eq.\eqref{KountertermCoupling} is the effective AdS radius of the vacuum
(maximally symmetric) solution about which the action is renormalized. In
other words, the renormalized action evaluated in that vacuum is zero, and for
solutions that are continuously connected to that vacuum (by the value of the
corresponding black hole charges) it measures the free energy with respect to
said vacuum. The action defined in eq.\eqref{KountertermLovelockAction} has a
finite value as well as a well-defined variational principle for a large class
of solutions including black holes with rotation and electromagnetic charges
\cite{Kofinas:2007ns}.

\subsection{The $P\left(\mathcal{F}\right)  $ formulation
for even-dimensional AAdS manifolds}
\label{P(F)_Bulk}

The renormalized volume of an even-dimensional
AAdS-Einstein manifold can be defined 
as a polynomial in totally antisymmetric contractions
of the   tensor \cite{Anastasiou:2018mfk}
\begin{equation}
\mathcal{F}_{\nu_{1}\nu_{2}}^{\mu_{1}\mu_{2}}=R_{\phantom{\mu_{1}\mu_{2}}\nu_{1}\nu_{2}}^{\mu_{1}%
\mu_{2}}+\frac{1}{\ell^{2}}\delta_{\nu_{1}\nu_{2}}^{\mu_{1}\mu_{2}}.
\label{AdS_curv}%
\end{equation}
It can be checked that this definition matches the standard definition of
renormalized volume
in $D=4$ (for generic Poincar\'e-Einstein manifolds) and $6$ (only for asymptotically conformally flat manifolds) as found in the mathematical literature \cite{Anderson2000L2CA,Graham:1999jg,Chang:2005ska,Alexakis:2010zz,Albin:2005qka}.
In this section, we find the analogous $P\left(  \mathcal{F}\right)  $
polynomial corresponding to the Kounterterm-renormalized bulk Lovelock action
in even-dimensional AAdS manifolds, in a similar fashion to ref.~\cite{Albin:2020lse}.

The Euler theorem states that for odd $d$
\begin{equation}%
{\displaystyle\int\limits_{M}}
E_{d+1}-%
{\displaystyle\int\limits_{\partial M}}
d^d x B_{d}=\left(  4\pi\right)  ^{\frac{d+1}{2}}\left(  \frac{d+1}{2}\right)
!\chi\left(  M\right)    \,,
\label{Euler-Theo}%
\end{equation}
where $E_{d+1}$ is the Euler density of the $\left(  d+1\right)
-$dimensional bulk, $B_{d}$ is the boundary Chern form defined
in eq.\eqref{Kounterterm} and $\chi\left(  M\right)  $ is the Euler
characteristic of the bulk manifold.  In this case we can rewrite the
renormalized Lovelock-AdS action  \eqref{KountertermLovelockAction} as
\begin{equation}
I_{L}^{ren}\left[  M\right]  =\frac{1}{16\pi G}%
{\displaystyle\int\limits_{M}}
\left(
{\displaystyle\sum\limits_{p=0}^{\frac{d-1}{2}}}
\alpha_{p}L_{2p}+c_{d} E_{d+1}\right)  +\tau_{d}\chi\left(  M\right)
,
\end{equation}
where%
\begin{equation}
\tau_{d}=-\frac{c_{d}}{16\pi G}\left(  4\pi\right)  ^{\frac{d+1}{2}}\left(
\frac{d+1}{2}\right)  ! \,. \label{TauInv}%
\end{equation}
Upon defining the following quantities as
\begin{align}
\alpha_{\frac{d+1}{2}}  &  =c_{d} \,, \qquad L_{d+1}   =E_{d+1} \,, 
\label{cd}
\end{align}
we have that
\begin{equation}
I_{L}^{ren}\left[  M\right]  =\frac{1}{16\pi G}%
{\displaystyle\int\limits_{M}}
{\displaystyle\sum\limits_{p=0}^{\frac{d+1}{2}}}
\alpha_{p}L_{2p}+\tau_{d}\chi\left(  M\right)  .
\end{equation}
One may rewrite the latter expression using eq.\eqref{AdS_curv} such that
\begin{align}
I  &  =16\pi G\left(  I_{L}^{ren}\left[  M\right]  -\tau_{d}\chi\left(
M\right)  \right) \nonumber\\
&  =%
{\displaystyle\int\limits_{M}}
d^{d+1}x\sqrt{-\mathcal{G}}%
{\displaystyle\sum\limits_{p=0}^{\frac{d+1}{2}}}
\frac{\alpha_{p}}{2^{p}}\delta_{\mu_{1}\cdots\mu_{2p}}^{\nu_{1}\cdots\nu_{2p}%
}\left(  \mathcal{F}_{\nu_{1}\nu_{2}}^{\mu_{1}\mu_{2}}-\frac{1}{\ell
_{\text{eff}}^{2}}\delta_{\nu_{1}\nu_{2}}^{\mu_{1}\mu_{2}}\right)
\cdots\left(  \mathcal{F}_{\nu_{2p-1}\nu_{2p}}^{\mu_{2p-1}\mu_{2p}}-\frac
{1}{\ell_{\text{eff}}^{2}}\delta_{\nu_{2p-1}\nu_{2p}}^{\mu_{2p-1}\mu_{2p}%
}\right)  \nonumber\\
&={\displaystyle\int\limits_{M}}
d^{d+1}x\sqrt{-\mathcal{G}}%
{\displaystyle\sum\limits_{p=0}^{\frac{d+1}{2}}}
{\displaystyle\sum\limits_{j=0}^{p}}
\frac{\left(  -1\right)  ^{p-j}p!\left(  d+1-2j\right)  !\alpha_{p}}%
{2^{j}j!\left(  p-j\right)  !\left(  d+1-2p\right)  !\ell_{\text{eff}%
}^{2\left(  p-j\right)  }}\delta_{\mu_{1}\cdots\mu_{2j}}^{\nu_{1}\cdots
\nu_{2j}}\mathcal{F}_{\nu_{1}\nu_{2}}^{\mu_{1}\mu_{2}}\cdots\mathcal{F}%
_{\nu_{2j-1}\nu_{2j}}^{\mu_{2j-1}\mu_{2j}} \,,
\label{act-IP}
\end{align}
upon  using the delta identities given in  
Appendix \ref{AppendixA}.

We now proceed to show that the coefficients of the $\mathcal{F}^{0}$ and
$\mathcal{F}^{1}$ terms are zero. The coefficient of the $\mathcal{F}^{0}$
term is given by
\begin{align}
c_{0} &= 
{\displaystyle\sum\limits_{p=0}^{\frac{d+1}{2}}}
\frac{\left(  -1\right)  ^{p}\left(  d+1\right)  !\alpha_{p}}{\left(
d+1-2p\right)  !}\left(  \frac{1}{\ell_{\text{eff}}^{2}}\right)  ^{p} \nonumber\\
&  =%
{\displaystyle\sum\limits_{p=0}^{\frac{d-1}{2}}}
\frac{\left(  -1\right)  ^{p}\left(  d+1\right)  !\alpha_{p}}{\left(
d+1-2p\right)  !}\left(  \frac{1}{\ell_{\text{eff}}^{2}}\right)  ^{p}-\left(
{\displaystyle\sum\limits_{p=0}^{\frac{d-1}{2}}}
\frac{\left(  -1\right)  ^{p}\left(  d+1\right)  !\alpha_{p}}{\left(
d+1-2p\right)  !}\left(  \frac{2p}{d+1}\right)  \left(  \frac{1}%
{\ell_{\text{eff}}^{2}}\right)  ^{p}\right) \nonumber\\
&  =
{\displaystyle\sum\limits_{p=0}^{\frac{d-1}{2}}}
\frac{\left(  -1\right)  ^{p}d!\alpha_{p}}{\left(  d-2p\right)  !}\left(
\frac{1}{\ell_{\text{eff}}^{2}}\right)  ^{p} \nonumber \\
&=-d\left(  d-1\right)
\Delta\left(  \ell_{\text{eff}}^{-2}\right)  =0 \,,
\label{c_0}
\end{align}
using eqs.\eqref{KountertermCoupling} and \eqref{cd}. The last line follows from the definition of (any one of) the effective AdS radii given by
the characteristic polynomial $\Delta\left(  \ell_{\text{eff}}^{-2}\right)
=0$ of the Lovelock theory as defined in  eq.\eqref{CharPoly}, noting that 
$\left\lfloor \frac{d}{2}\right\rfloor =\left\lfloor \frac{d-1}{2}\right\rfloor $ for odd $d$.
 Turning to the coefficient of the $\mathcal{F}^{1}$ term, we have 
\begin{align}
c_{1} &= 
{\displaystyle\sum\limits_{p=1}^{\frac{d+1}{2}}}
\frac{\left(  -1\right)  ^{p-1}p\left(  d-1\right)  !\alpha_{p}}{2\left(
d+1-2p\right)  !\ell_{\text{eff}}^{2\left(  p-1\right)  }} \nonumber\\
&={\displaystyle\sum\limits_{p=1}^{\frac{d-1}{2}}}
\left(  \frac{\left(  -1\right)  ^{p-1}p\left(  d-1\right)  !\alpha_{p}%
}{2\left(  d+1-2p\right)  !\ell_{\text{eff}}^{2\left(  p-1\right)  }}%
+\frac{\left(  -1\right)  ^{d}\left(  -1\right)  ^{p-1}p\left(  d-1\right)
!\alpha_{p}}{2\left(  d+1-2p\right)  !\ell_{\text{eff}}^{2\left(  p-1\right)
}}\right)  \nonumber\\
&=0 \,,
 \label{c_1}
\end{align}
considering again eqs.\eqref{KountertermCoupling} and \eqref{cd} and noting that $\left(  -1\right)  ^{d}=-1$.

Thus to lowest order in the AdS curvature, the integrand $P\left(  \mathcal{F}\right)
$ in eq.\eqref{act-IP} is of quadratic and higher order in  $\mathcal{F}$, which (as discussed in Appendix \ref{AppendixB})
also implies that the Noether prepotential is proportional to $\mathcal{F}$ at
the normalizable order (assuming a non-degenerate theory). We then have that
\begin{equation}
I_{L}^{ren}\left[  M\right]  =\frac{1}{16\pi G}%
{\displaystyle\int\limits_{M}}
d^{d+1}x\sqrt{-\mathcal{G}}P_{\left(  d+1\right)  ,\left\{  \alpha
_{p}\right\}  }\left(  \mathcal{F}\right)  +\tau_{d}\chi\left(  M\right)  ,
\end{equation}
for the Kounterterm-renormalized Lovelock-AdS action for odd $d$, where
\begin{equation}
P_{\left(  d+1\right)  ,\left\{  \alpha_{p}\right\}  }\left(  \mathcal{F}%
\right)  =%
{\displaystyle\sum\limits_{p=2}^{\frac{d+1}{2}}}
{\displaystyle\sum\limits_{j=2}^{p}}
\frac{\left(  -1\right)  ^{p-j}p!\left(  d+1-2j\right)  !\alpha_{p}}%
{2^{j}j!\left(  p-j\right)  !\left(  d+1-2p\right)  !\ell_{\text{eff}%
}^{2\left(  p-j\right)  }}\delta_{\mu_{1}\cdots\mu_{2j}}^{\nu_{1}\cdots
\nu_{2j}}\mathcal{F}_{\nu_{1}\nu_{2}}^{\mu_{1}\mu_{2}}\cdots\mathcal{F}%
_{\nu_{2j-1}\nu_{2j}}^{\mu_{2j-1}\mu_{2j}}, \label{BulkP(F)}%
\end{equation}
and with $\tau_{d}$ given in terms of the Euler characteristic of the bulk
manifold $M_{d+1}$ from \eqref{TauInv}. We note that sometimes it is more convenient to write 
\begin{equation}
P_{\left(  d+1\right)  ,\left\{  \alpha_{p}\right\}  }\left(  \mathcal{F}%
\right)  =%
{\displaystyle\sum\limits_{j=2}^{\frac{d+1}{2}}}
c_{j}\delta_{\mu_{1}\cdots\mu_{2j}}^{\nu_{1}\cdots\nu_{2j}}\mathcal{F}%
_{\nu_{1}\nu_{2}}^{\mu_{1}\mu_{2}}\cdots\mathcal{F}_{\nu_{2j-1}\nu_{2j}}%
^{\mu_{2j-1}\mu_{2j}}, 
\label{P(F)BulkFixedJ}%
\end{equation}
in order to immediately identify the coefficient $c_{j}$ of the $\mathcal{F}%
^{j}$ term as%
\begin{equation}
c_{j}=%
{\displaystyle\sum\limits_{p=j}^{\frac{d+1}{2}}}
\frac{\left(  -1\right)  ^{p-j}p!\left(  d+1-2j\right)  !\alpha_{p}}%
{2^{j}j!\left(  p-j\right)  !\left(  d+1-2p\right)  !\ell_{\text{eff}%
}^{2\left(  p-j\right)  }}. \label{Cj}%
\end{equation}
In the Einstein-AdS case, where  $\alpha_{0}$ and $\alpha_{1}$ are given in  \eqref{Einstein-AdS}, with $\alpha_{p>1}=0$, 
the expression \eqref{P(F)BulkFixedJ} matches that obtained previously
for the definition of renormalized volume \cite{Anastasiou:2018mfk}.

We show in Appendix \ref{AppendixC} that 
\begin{equation}
c_{k+1}=%
{\displaystyle\sum\limits_{i=1}^{k}}
p_{\left(  k+1,i\right)  }\Delta^{\left(  i\right)  } \,,
\label{GeneralDegeneracy}%
\end{equation}
for some coefficients $p_{\left(  k+1,i\right)  }$ (given in eq.\eqref{P_ij}), in agreement with the
previous results for Lovelock gravity with $k$-fold degenerate vacua \cite{Arenas-Henriquez:2019rph}. 
We can explicitly write the first three factors as

\begin{align}
c_{2}  & =\frac{\ell_{\text{eff}}^{2}}{2^{3}2!\left(  d-2\right)  }\left(
\Delta^{\left(  1\right)  }\right)  ,\nonumber\\
c_{3}  & =-\frac{\ell_{\text{eff}}^{4}}{2^{5}3!\left(  d-2\right)  \left(
d-4\right)  }\left(  \Delta^{\left(  1\right)  }+\frac{4}{\left(  d-3\right)
\ell_{\text{eff}}^{2}}\Delta^{\left(  2\right)  }\right)  ,\nonumber\\
c_{4}  & =\frac{\ell_{\text{eff}}^{6}}{2^{7}4!\left(  d-2\right)  \left(
d-4\right)  \left(  d-6\right)  }\left(  \Delta^{\left(  1\right)  }+\frac
{4}{\left(  d-3\right)  \ell_{\text{eff}}^{2}}\Delta^{\left(  2\right)
}+\frac{24}{\left(  d-3\right)  \left(  d-5\right)  \ell_{\text{eff}}^{4}%
}\Delta^{\left(  3\right)  }\right)  .
\end{align}
Also, the generic factor is given by%
\begin{align}
c_{i}  & =\frac{\left(  -1\right)  ^{i}\ell_{\text{eff}}^{2i-2}}%
{2^{2i-1}i!\left(  d-2\right)  \left(  d-4\right)  \cdots\left(
d+2-2i\right)  }\left(  \Delta^{\left(  1\right)  }+\frac{4}{\left(
d-3\right)  \ell_{\text{eff}}^{2}}\Delta^{\left(  2\right)  }\right.
\nonumber\\
& \left.  +\frac{24}{\left(  d-3\right)  \left(  d-5\right)  \ell_{\text{eff}%
}^{4}}\Delta^{\left(  3\right)  }+\cdots+\frac{2^{\left(  i-1\right)  }%
i!}{\ell_{\text{eff}}^{2\left(  i-2\right)  }\left(  d-3\right)  \left(
d-5\right)  \cdots\left(  d+3-2i\right)  }\Delta^{\left(  i-1\right)
}\right).\label{GenericCoeff}
\end{align}
Taking advantage of eq.\eqref{GeneralDegeneracy}  we can write 
\begin{equation}
P_{\left(  d+1\right)  ,\left\{  \alpha_{p}\right\}  }\left(  \mathcal{F}%
\right)  =%
{\displaystyle\sum\limits_{j=2}^{\frac{d+1}{2}}}
{\displaystyle\sum\limits_{i=1}^{j-1}}
\left(  p_{\left(  j,i\right)  }\Delta^{\left(  i\right)  }\right)
\delta_{\mu_{1}\cdots\mu_{2j}}^{\nu_{1}\cdots\nu_{2j}}\mathcal{F}_{\nu_{1}%
\nu_{2}}^{\mu_{1}\mu_{2}}\cdots\mathcal{F}_{\nu_{2j-1}\nu_{2j}}^{\mu_{2j-1}%
\mu_{2j}}
\end{equation}
thereby relating  $P\left(  \mathcal{F}\right)$ to the degeneracy conditions \eqref{k-thDeg}. 

For a $k$-fold degenerate vacuum, all the degeneracy conditions up to
(and including)   $\Delta^{\left(  k\right)  }$ are zero as seen from their
definition in eq.\eqref{k-thDeg}. Thus, the lowest order in $\mathcal{F}$ of
the $P\left(  \mathcal{F}\right)  $ that encodes the renormalized action of a
$k$-degenerate theory (normalized with respect to the $k$-degenerate vacuum)
is $\mathcal{F}^{k+2}$.  Furthermore   the $P_{\left(  d+1\right)  ,\left\{
\alpha_{p}\right\}  }\left(  \mathcal{F}\right)  $ can always be written as%
\begin{equation}
P_{\left(  d+1\right)  ,\left\{  \alpha_{p}\right\}  }\left(  \mathcal{F}%
\right)  =c_{k+2}\delta_{\mu_{1}\cdots\mu_{2k+4}}^{\nu_{1}\cdots\nu_{2k+3}%
}\mathcal{F}_{\nu_{1}\nu_{2}}^{\mu_{1}\mu_{2}}\cdots\mathcal{F}_{\nu_{2k+3}%
\nu_{2k+4}}^{\mu_{2k+3}\mu_{2k+4}}+%
{\displaystyle\sum\limits_{j=k+3}^{\frac{d+1}{2}}}
c_{j}\delta_{\mu_{1}\cdots\mu_{2j}}^{\nu_{1}\cdots\nu_{2j}}\mathcal{F}%
_{\nu_{1}\nu_{2}}^{\mu_{1}\mu_{2}}\cdots\mathcal{F}_{\nu_{2j-1}\nu_{2j}}%
^{\mu_{2j-1}\mu_{2j}},
\end{equation}
where
\begin{equation}
c_{k+2}=p_{\left(  k+2,k+1\right)  }\Delta^{\left(  k+1\right)  }
\end{equation}
is proportional to the $\Delta^{\left(  k+1\right)  }$ degeneracy condition
and $p_{\left(  k+2,k+1\right)  }$ is given by%
\begin{equation}
p_{\left(  k+2,k+1\right)  }=\frac{\ell_{\text{eff}}^{2}\left(  -1\right)
^{k-1}\left(  d-1-2k\right)  !}{2^{k+2}\left(  k+1\right)  \left(  d-2\right)
!} \,,\label{DegP}%
\end{equation}
in a $k$-degenerate theory.

Thus, the rewriting of the coefficients in terms of the degeneracy conditions indicates that the Noether prepotential of a $k$-degenerate theory is
of order $\mathcal{F}^{k+1}$ at the normalizable order, in accordance to ref.\cite{Arenas-Henriquez:2019rph}. For more information, see Appendix \ref{AppendixD}.

In the next section, we consider the Kounterterm-renormalized Lovelock-AdS
action discussed here, together with the Lewkowycz-Maldacena (LM) procedure \cite{Lewkowycz:2013nqa}, in order to compute the renormalized HEE by evaluating the action on the replica orbifold.

\section{Renormalized Lovelock-AdS action on the
replica orbifold and HEE}
\label{ActionOnOrbifold}

Our next task is to evaluate the renormalized (Euclidean) Lovelock-AdS action on the
conically-singular orbifold $M_{d+1}^{\left(  \alpha\right)  }$. The Lovelock densities
evaluated on the replica orbifold decompose into the sum of the regular bulk
part and a co-dimension 2 Lovelock density localized at the extremal surface $\Sigma$
which corresponds to the fixed-point set of the replica symmetry \cite{Fursaev:2013fta,Anastasiou:2019ldc,Kastikainen:2020auf}.  
In particular
\begin{equation}%
{\displaystyle\int\limits_{M_{d+1}^{\left(  \alpha\right)  }}}
d^{d+1}x\sqrt{\mathcal{G}}L_{2p}^{\left(  \alpha\right)  }=%
{\displaystyle\int\limits_{M_{d+1}}}
d^{d+1}x\sqrt{\mathcal{G}}L_{2p}^{\left(  r\right)  }+4\pi p\left(
1-\alpha\right)
{\displaystyle\int\limits_{\Sigma}}
d^{d-1}y\sqrt{\gamma}L_{2p-2}, \label{Lovelock_Decomp}%
\end{equation}
where $L_{2p-2}$ is an intrinsic Lovelock density evaluated on the co-dimension 2
surface $\Sigma$ with induced metric $\gamma$. Therefore, the bulk
part of the Lovelock action, when evaluated on the orbifold, decomposes as%
\begin{align}
I_{L}^{bulk}\left[  M_{d+1}^{\left(  \alpha\right)  }\right]   &  =\frac
{1}{16\pi G}%
{\displaystyle\sum\limits_{p=0}^{\left\lfloor \frac{d}{2}\right\rfloor }}
\alpha_{p}%
{\displaystyle\int\limits_{M_{d+1}}}
d^{d+1}x\sqrt{\mathcal{G}}L_{2p}^{\left(  r\right)  }+\left(  1-\alpha\right)
S_{JM}\left[  \Sigma\right]  ,\nonumber\\
S_{JM}\left[  \Sigma\right]   &  =\frac{1}{4G}%
{\displaystyle\sum\limits_{p=1}^{\left\lfloor \frac{d}{2}\right\rfloor }}
\alpha_{p}p%
{\displaystyle\int\limits_{\Sigma}}
d^{d-1}y\sqrt{\gamma}L_{2p-2} \,,
\label{ICod2Bulk}%
\end{align}
where  $S_{JM}\left[  \Sigma\right]  $ is precisely the
Jacobson-Myers (JM) functional  \cite{Hung:2011xb,deBoer:2011wk}. Thus, in
analogy with the Einstein-AdS case, the action on the orbifold is interpreted
as the bulk contribution plus the action of a brane with tension
$T=\frac{\left(  1-\alpha\right)  }{4G}$, but coupled to the bulk geometry
through the JM functional, which has the form of a co-dimension 2 Lovelock Lagrangian
evaluated on the intrinsic metric of the brane.

To evaluate the boundary Kounterterm on the orbifold, we consider the
self-replicating property of the $B_{d}$ (defined in eq.\eqref{Kounterterm}),
for both the odd and even $d$ cases  \cite{Anastasiou:2019ldc}.
In particular, one writes
\begin{equation}%
{\displaystyle\int\limits_{\partial M_{d+1}^{\left(  \alpha\right)  }}}
d^d x B_{d}^{\left(  \alpha\right)  }=%
{\displaystyle\int\limits_{\partial M_{d+1}^{\left(  \alpha\right)  }}}
d^d x B_{d}^{\left(  r\right)  }+4\pi\left\lfloor \frac{d+1}{2}\right\rfloor \left(
1-\alpha\right)
{\displaystyle\int\limits_{\partial\Sigma}}
d^{d-2} y B_{d-2}.
\end{equation}
Then, the evaluation of the counterterm results in%
\begin{equation}
I_{L}^{B_{d}}\left[  M_{d+1}^{\left(  \alpha\right)  }\right]  =\frac{c_{d}%
}{16\pi G}%
{\displaystyle\int\limits_{\partial M_{d+1}^{\left(  \alpha\right)  }}}
d^d x B_{d}^{\left(  r\right)  }+\frac{c_{d}\left\lfloor \frac{d+1}{2}\right\rfloor
}{4G}\left(  1-\alpha\right)
{\displaystyle\int\limits_{\partial\Sigma}}
d^{d-2} y B_{d-2}, \label{ICod2Boundary}%
\end{equation}
where the coupling $c_{d}$ is given in eq.\eqref{KountertermCoupling}.
Combining eqs.\eqref{ICod2Bulk} and \eqref{ICod2Boundary}, we have 
\begin{align}
I_{L}^{ren}\left[  M_{d+1}^{\left(  \alpha\right)  }\right]   &  = 
\frac{1}{16\pi G} \left(
{\displaystyle\sum\limits_{p=0}^{\left\lfloor \frac{d}{2}\right\rfloor }}
\alpha_{p} 
{\displaystyle\int\limits_{M_{d+1}}}
d^{d+1}x\sqrt{G}L_{2p}^{\left(  r\right)  }+c_{d}%
{\displaystyle\int\limits_{\partial M_{d+1}^{\left(  \alpha\right)  }}}
d^d x B_{d}^{\left(  r\right)  }\right)   \nonumber\\
& \qquad +     \left(  1-\alpha\right)  \left(  S_{JM}\left[  \Sigma\right]
+\frac{c_{d}\left\lfloor \frac{d+1}{2}\right\rfloor }{4G} 
{\displaystyle\int\limits_{\partial\Sigma}}
d^{d-2 y} B_{d-2}\right)  \,,
\label{Iren_alpha}
\end{align}
what is the renormalized Lovelock-AdS action evaluated on the replica
orbifold.

From this expression, we compute the renormalized HEE using the replica formula of the LM prescription, given in
eq.\eqref{Replica_EE}. Starting from the action on $M_{d+1}^{\left(
\alpha\right)  }$, given by eq.\eqref{Iren_alpha}, we obtain
\begin{equation}
S_{EE}^{ren}=-\left.  \partial_{\alpha}I_{E}^{ren}\left[  M_{d+1}^{\left(
\alpha\right)  }\right]  \right\vert _{\alpha=1}=S_{JM}\left[  \Sigma\right]
+\frac{c_{d}\left\lfloor \frac{d+1}{2}\right\rfloor }{4G}
{\displaystyle\int\limits_{\partial\Sigma}} d^{d-2} y B_{d-2} \equiv S_{JM}^{ren}\left[  \Sigma\right] \,,
\label{RenHEELovelock}
\end{equation}
what defines the renormalized JM functional. Note that $S_{JM}\left[  \Sigma\right]  $ is the JM functional evaluated on the
extremal surface $\Sigma$ that minimizes it. 
We then have that the renormalized HEE, which directly corresponds to its
universal part, is given by the renormalized JM functional.

The extremal surface  minimizing this new functional is not affected by the counterterms, since it's only a boundary term that does not affect the dynamics. In the tensionless limit ($\alpha\rightarrow1$), there
is no back-reaction of the extremal surface on the bulk geometry, as seen by
the fact that the contribution to the action from the surface (given by the
$JM$ functional in eq.\eqref{Iren_alpha}) vanishes. Thus, the extremal surface
is found by finding determined by the global minimum of $S_{JM}$ by itself. The resulting
Euler-Lagrange equations, given in ref. \cite{Bhattacharyya:2014yga}, obtain the form below
\begin{equation}
E_{JM}=2\mathcal{K}_{\beta}^{\alpha}\left(
{\displaystyle\sum\limits_{p=0}^{\left\lfloor \frac{d-2}{2}\right\rfloor }}
\frac{\alpha_{\left(  p+1\right)  }\left(  p+1\right)  }{2^{p+1}}%
\delta_{\alpha\alpha_{1}\cdots\alpha_{2p}}^{\beta\beta_{1}\cdots\beta_{2p}%
}\widehat{\mathcal{R}}_{\phantom{\alpha_{1}\alpha_{2}}\beta_{1}\beta_{2}}^{\alpha_{1}\alpha_{2}}\cdots\widehat{\mathcal{R}}%
_{\phantom{\alpha_{2p-1}\alpha_{2p}}\beta_{2p-1}\beta_{2p}}^{\alpha_{2p-1}\alpha_{2p}}\right)  =0, 
\label{JMEOM}%
\end{equation}
where $\mathcal{K}_{\beta}^{\alpha}$ is the extrinsic curvature of the surface
with respect to the normal direction that is not along the time coordinate,
and $\widehat{\mathcal{R}}_{\phantom{\alpha_{1}\alpha_{2}}\beta_{1}\beta_{2}}^{\alpha_{1}\alpha_{2}}$ is the intrinsic
Riemann curvature of the surface. 

The form of the EOM in eq.\eqref{JMEOM} is very similar to that in eq.\eqref{LovelockEOM} for Lovelock gravity, but with a very important difference. The co-dimension 2 Lanczos-Lovelock tensor is contracted with the extrinsic
curvature of the minimal surface, which comes from the variation of the
induced metric. Following the same factorization procedure discussed in  
Appendix \ref{AppendixB}, the eq.\eqref{JMEOM} can be rewritten as
\begin{align}
E_{JM} & =\mathcal{K}_{\beta}^{\alpha}\frac{\alpha_{N}N}{2^{N-1}}\delta
_{\alpha\alpha_{1}\cdots\alpha_{2N-2}}^{\beta\beta_{1}\cdots\beta_{2N-2}%
}\left(  \widehat{\mathcal{R}}_{\phantom{\alpha_{1}\alpha_{2}}\beta_{1}\beta_{2}}^{\alpha_{1}\alpha_{2}}%
+\lambda_{\left(  1\right)  }\delta_{\beta_{1}\beta_{2}}^{\alpha_{1}\alpha
_{2}}\right) \left(  \widehat{\mathcal{R}}_{\phantom{\alpha_{3}\alpha_{4}}\beta_{3}\beta_{4}}^{\alpha_{3}\alpha_{4}}%
+\lambda_{\left(  2\right)  }\delta_{\beta_{3}\beta_{4}}^{\alpha_{3}\alpha
_{3}}\right)  \nonumber\\ 
&\qquad \qquad\qquad
 \cdots\left(  \widehat{\mathcal{R}}_{\phantom{\alpha_{2N-3}\alpha_{2N-2}}\beta_{2N-3}\beta_{2N-2}}%
^{\alpha_{2N-3}\alpha_{2N-2}}+\lambda_{\left(  N-1\right)  }\delta
_{\beta_{2N-3}\beta_{2N-2}}^{\alpha_{2N-3}\alpha_{2N-2}}\right)  =0,
\end{align}
where $N\leq\left\lfloor \frac{d}{2}\right\rfloor $ and $\lambda_{\left(
i\right)  }$ are the roots of the polynomial
\begin{equation}
D\left(  \lambda\right)  \overset{\text{def}}{=}%
{\displaystyle\sum\limits_{p=0}^{N-1}}
\frac{\left(  -1\right)  ^{p}\left(  d-2\right)  !\alpha_{\left(  p+1\right)
}\left(  p+1\right)  }{\left(  d-2p-2\right)  !}\lambda^{p}=0.
\end{equation}
Note that the $\lambda_{\left(  i\right)  }$ solutions are different from the
roots of $\Delta\left(  \ell_{\text{eff}}^{-2}\right)  $,  which define the
vacua of the gravity theory.

Finally, it is easy to check that for a ball-shaped entangling region in a
pure AdS bulk (dual to the ground state of a CFT in Minkowski spacetime), the
extremal surface of the JM functional is the same as the RT minimal surface
(i.e., a spherical hemisphere). We show this in detail in  Appendix \ref{AppendixE}.

Having obtained the renormalized HEE functional for Lovelock-AdS gravity, we
proceed in the next section to show the explicit cancellation of leading order
and next-to-leading order divergencies of said renormalized HEE.

\section{Renormalized HEE divergence
cancelation in Lovelock-AdS}
\label{DivergenceCancellation}

In what follows, we check the cancellation of divergences in the renormalized
HEE $S_{EE}^{\text{ren}}$ given in  eq.\eqref{RenHEELovelock}, for both even
and odd dimensional manifolds, up to the next-to-leading order in the
holographic radial coordinate $\rho$. The $S_{EE}^{\text{ren}}$ consists of
the sum of the JM functional  (\ref{ICod2Bulk}) and the co-dimension 2 Kounterterm,
both evaluated on the extremal surface $\Sigma$. 
The induced metric $\gamma_{\alpha\beta}$ of this
extremal surface $\Sigma$ has an FG-like expansion
\begin{align}
ds_{\gamma}^{2}  &  =\gamma_{\alpha\beta}dy^{\alpha}dy^{\beta}=N^{2}\left(
\rho\right)  d\rho^{2}+\widetilde{\gamma}_{ab}dy^{a}dy^{b},~\nonumber\\
N^{2}\left(  \rho\right)   &  =\frac{\ell^{2}}{4\rho^{2}}\left(  1+\frac
{\rho\ell^{2}\kappa_{a}^{\hat{\imath}a}\kappa_{b}^{\hat{\imath}b}}{\left(
d-2\right)  ^{2}}+\ldots\right) \,, \qquad \widetilde{\gamma}_{ab}   =\frac{\sigma_{ab}}{\rho} \,, \nonumber\\
~\sigma_{ab} & =\sigma
_{ab}^{\left(  0\right)  }+\rho\sigma_{ab}^{\left(  2\right)  }+\ldots \,,
\qquad 
\sigma_{ab}^{\left(  2\right)  }    =-\ell^{2}S_{ab}-\frac{\ell^{2}}{\left(
d-2\right)  }\kappa_{c}^{\hat{\imath}c}\kappa_{ab}^{\hat{\imath}},
\label{FG-likeMetric}%
\end{align}
where  $\kappa_{ab}^{\hat{\imath}}$ is the extrinsic
curvature of the boundary of the extremal surface $\partial\Sigma$ along the
$\hat{\imath}$ direction (normal to the radial coordinate $\rho$) and $\widetilde{\gamma}_{ab}$ is the induced metric on
$\partial\Sigma$   \cite{Anastasiou:2019ldc,Schwimmer:2008yh,Hung:2011nu}. 
Furthermore,  $\widetilde{\gamma}_{ab}$ has an FG-like expansion
whose leading and next-to-leading order coefficients are given by
$\sigma^{\left(  0\right)  }$ and $\sigma^{\left(  2\right)  }$, where
$\sigma^{\left(  2\right)  }$ depends on the Schouten tensor $S_{ab}$ of the
CFT metric $g_{ij}^{\left(  0\right)  }$ evaluated with the indices on
$\partial\Sigma$. Note that the Greek letters denote directions along the
world-volume of $\Sigma$, whereas the lower case Latin letters denote
directions along $\partial\Sigma$. 

Decomposing the Riemann curvature tensor along and orthogonal to the holographic radial foliation,
we obtain
\begin{align}
\widehat{\mathcal{R}}_{\phantom{\rho a}\rho b}^{\rho a}  &  =\frac{1}{N}\partial_{\rho}%
k_{b}^{a}-k_{c}^{a}k_{b}^{c},\nonumber\\
\widehat{\mathcal{R}}_{\phantom{\rho a} bc}^{a\rho}  &  =\frac{2}{N}\nabla_{\lbrack b}%
k_{c]}^{a} \qquad ~\widehat{\mathcal{R}}_{\phantom{\rho a} c\rho}^{ab}=2N\nabla^{\lbrack a}k_{c}%
^{b]},\nonumber\\
\widehat{\mathcal{R}}_{\phantom{ba}cd}^{ab}  &  =\mathcal{R}_{\phantom{ba}cd}^{ab}-2k_{c}^{[a}%
k_{d}^{b]}, \label{RiemannDecomp}%
\end{align}
 using   the Gauss-Codazzi relations \cite{Toolkit}, 
where $k_{b}^{a}$ is the extrinsic curvature of $\partial\Sigma$ along the
radial direction $\partial_{\rho}$, and $\nabla_{a}$ is the covariant
derivative with respect to $\widetilde{\gamma}_{ab}$. In order to avoid
confusion, we denote the Riemann tensor of $\gamma_{\alpha\beta}$ with a hat
$\left(  \widehat{\mathcal{R}}_{\phantom{\alpha_{1}\alpha_{2}}\beta_{1}\beta_{2}}^{\alpha_{1}\alpha_{2}}\right)  $, and that of $\widetilde{\gamma}_{ab}$ without a hat $\left(
\mathcal{R}_{\phantom{a_{1}a_{2}}b_{1}b_{2}}^{a_{1}a_{2}}\right)  $. Finally,  
the FG-like expansion of the co-dimension 3 curvatures (Riemannian
curvature and extrinsic curvature along $\partial_{\rho}$) at the boundary
$\partial\Sigma$ is given by
\begin{align}
\mathcal{R}_{\phantom{a_{1}a_{2}}b_{1}b_{2}}^{a_{1}a_{2}}  &  =\rho \mathcal{R}_{\phantom{a_{1}a_{2}}b_{1}b_{2}}^{a_{1}a_{2}}\left[  \sigma\right]  =\rho\left(  \mathcal{R}^{\left(
0\right)  }\right)_{\phantom{a_{1}a_{2}}b_{1}b_{2}}^{a_{1}a_{2}}+\ldots,\text{ }k_{b}%
^{a}=\left(  k^{\left(  0\right)  }\right)  _{b}^{a}+\rho\left(  k^{\left(
2\right)  }\right)  _{b}^{a}+\ldots,\nonumber\\
\left(  k^{\left(  0\right)  }\right)  _{b}^{a}  &  =\frac{1}{\ell
_{\text{eff}}}\delta_{b}^{a},\text{ }\left(  k^{\left(  2\right)  }\right)
_{b}^{a}=-\frac{1}{\ell_{\text{eff}}}\left[  \left(  \sigma^{\left(  2\right)
}\right)  _{b}^{a}+\frac{\ell_{\text{eff}}^{2}\kappa_{c}^{\hat{\imath}c}%
\kappa_{d}^{\hat{\imath}d}}{2(d-2)}\delta_{b}^{a}\right]  ,
\label{FG-likeTensors}%
\end{align}
where we note that the $\left(  0\right)  $-quantities are computed with
respect to the intrinsic metric $\sigma_{ab}^{\left(  0\right)  }$ on the
entangling surface in the CFT \cite{Anastasiou:2019ldc}. Although the FG-like expansions in eqs.\eqref{FG-likeMetric} and \eqref{FG-likeTensors} were used in ref.\cite{Anastasiou:2019ldc}, 
they are also valid for Lovelock-AdS.

We now define%
\begin{align}
4GS_{EE}^{\text{ren}}  &  =I_{JM}+I_{KT},\nonumber\\
I_{JM}  &  =%
{\displaystyle\int\limits_{\Sigma}}
d^{d-1}y\sqrt{\gamma}%
{\displaystyle\sum\limits_{p=0}^{\left\lfloor \frac{d-2}{2}\right\rfloor }}
\alpha_{\left(  p+1\right)  }\left(  p+1\right)  L_{2p} \,, \qquad I_{KT}    =c_{d}\left\lfloor \frac{d+1}{2}\right\rfloor \int_{\partial\Sigma
} d^{d-2} yB_{d-2}  \,,
\label{I_definitions}\\
L_{2p}  &  =\frac{1}{2^{p}}\delta_{\alpha_{1}\cdots\alpha_{2p}}^{\beta
_{1}\cdots\beta_{2p}}  \widehat{\mathcal{R}}_{\phantom{\alpha_{1}\alpha_{2}}\beta_{1}\beta_{2}}^{\alpha_{1}\alpha_{2}} \cdots\widehat{\mathcal{R}}_{\phantom{\alpha_{2p-1}\alpha_{2p}}\beta_{2p-1}\beta_{2p}}%
^{\alpha_{2p-1}\alpha_{2p}} \,, \nonumber
\end{align}
with $c_{d}$ given in eq.(\ref{KountertermCoupling}), and proceed with the
computation of $I_{JM}$ and $I_{KT}$.

In order to isolate the divergences coming from the bulk term at $\Sigma$, we
first expand the $I_{JM}$ of eq.(\ref{I_definitions}) in the radial foliation.
Using the antisymmetry of the generalized Kronecker delta, $I_{JM}$ can
be expanded as%
\begin{align}
I_{JM}  &  =I_{JM}^{\left(  1\right)  }+I_{JM}^{\left(  2\right)  }%
+I_{JM}^{\left(  3\right)  },\nonumber\\
I_{JM}^{\left(  1\right)  }  &  =\int_{\Sigma}d^{d-1}y\sqrt{\gamma}%
\sum\limits_{p=1}^{\left\lfloor \frac{d-2}{2}\right\rfloor }\frac{\left(
p+1\right)  p\alpha_{\left(  p+1\right)  }}{2^{p-2}}\delta_{b_{1}\ldots
b_{2p-1}}^{a_{1}\ldots a_{2p-1}}
\widehat{\mathcal{R}}_{\phantom{\rho b_{1}}\rho a_{1}}^{\rho b_{1}}\widehat{\mathcal{R}}_{\phantom{b_{2}b_{3}}a_{2}a_{3}}^{b_{2}b_{3}}\cdots\widehat
{\mathcal{R}}_{\phantom{b_{2p-2}b_{2p-1}}a_{2p-2}a_{2p-1}}^{b_{2p-2}b_{2p-1}},\nonumber\\
I_{JM}^{\left(  2\right)  }  &  =\int_{\Sigma}d^{d-1}y\sqrt{\gamma}%
\sum\limits_{p=2}^{\left\lfloor \frac{d-2}{2}\right\rfloor }\frac{\left(
p+1\right)  \left(  p-1\right)  p\alpha_{\left(  p+1\right)  }}{2^{p-2}}%
\delta_{b_{1}\ldots b_{2p-1}}^{a_{1}\ldots a_{2p-1}}\widehat{\mathcal{R}
}_{\phantom{\rho b_{1}} a_{1}a_{2}}^{\rho b_{1}}\widehat{\mathcal{R}}_{\phantom{b_{2}b_{3}}\rho a_{3}}^{b_{2}b_{3}%
}\widehat{\mathcal{R}}_{\phantom{b_{4}b_{5}}a_{4}a_{5}}^{b_{4}b_{5}}\cdots
{\mathcal{R}}_{\phantom{b_{2p-2}b_{2p-1}}a_{2p-2}a_{2p-1}}^{b_{2p-2}b_{2p-1}},\nonumber\\
I_{JM}^{\left(  3\right)  }  &  =\int_{\Sigma}d^{d-1}y\sqrt{\gamma}%
\sum\limits_{p=0}^{\left\lfloor \frac{d-2}{2}\right\rfloor }\frac{\left(
p+1\right)  \alpha_{\left(  p+1\right)  }}{2^{p}}\delta_{b_{1}\ldots b_{2p}%
}^{a_{1}\ldots a_{2p}}\widehat{\mathcal{R}}_{\phantom{b_{1}b_{2}}a_{1}a_{2}}^{b_{1}b_{2}}%
\cdots\widehat{\mathcal{R}}_{\phantom{b_{2p-1}b_{2p}}a_{2p-1}a_{2p}}^{b_{2p-1}b_{2p}},
\label{I_JM_decomp}%
\end{align}
where we have separated the indices of all the possible terms into those
corresponding to coordinates along the worldvolume of $\partial\Sigma$
(denoted by the Latin lowercase letters $a$, $b$) and the radial coordinate
$\rho$.  Each of these terms can be simplified using eqs.(\ref{FG-likeMetric}-\ref{FG-likeTensors}); as shown in  Appendix \ref{AppendixF}, the result is
\begin{align}
I_{JM}  &  =%
{\displaystyle\int\limits_{\Sigma}}
d^{d-1}y\sqrt{\gamma}\sum\limits_{p=0}^{\left\lfloor \frac{d-2}{2}%
\right\rfloor }\frac{\left(  -1\right)  ^{p}\left(  p+1\right)  \left(
d-1\right)  !\alpha_{\left(  p+1\right)  }}{\ell_{\text{eff}}^{2p}\left(
d-1-2p\right)  !}\Biggl(  1 \nonumber\\
&  \left.  -\frac{\rho p}{\left(  d-2\right)  \left(  d-1\right)  }\left[
2(d-3)\left(  tr\left[  \sigma^{\left(  2\right)  }\right]  +\frac
{\ell_{\text{eff}}^{2}\kappa_{a}^{\left(  i\right)  a}\kappa_{b}^{\left(
i\right)  b}}{2(d-2)}\right)  +\ell_{\text{eff}}^{2}\mathcal{R}\left[
\sigma\right]  \right]  \right)  +\text{h.o.} \,. \label{I_JM_sum}%
\end{align}
Employing the FG-like expansion 
\begin{equation}
\sqrt{\gamma}=\frac{\ell_{\text{eff}}\sqrt{\sigma^{\left(  0\right)  }}}%
{2\rho^{\frac{d}{2}}}\left(  1+\frac{\rho}{2}\left(  tr\left[  \sigma^{\left(
2\right)  }\right]  +\frac{\ell_{\text{eff}}^{2}\kappa_{a}^{\left(  i\right)
a}\kappa_{b}^{\left(  i\right)  b}}{(d-2)^{2}}\right)  +\mathcal{O}\left(
\rho^{2}\right)  \right) \,,
\end{equation}
of $\sqrt{\gamma}$  
and decomposing the volume element on $\Sigma$ into its radial and transverse
components, we obtain
\begin{align}
I_{JM}  &  =C_{1}+%
{\displaystyle\int\limits_{\partial\Sigma}}
\frac{d^{d-2}y\ell\sqrt{\sigma^{\left(  0\right)  }}}{\left(  d-2\right)
\epsilon^{\frac{d-2}{2}}}%
{\displaystyle\sum\limits_{p=0}^{\left\lfloor \frac{d-2}{2}\right\rfloor }}
\frac{\left(  -1\right)  ^{p}\left(  p+1\right)  \left(  d-1\right)
!\alpha_{\left(  p+1\right)  }}{\left(  d-1-2p\right)  !\ell_{\text{eff}}%
^{2p}}   \nonumber\\
&\qquad\qquad \times \left[1 +  \frac{\epsilon}{\left(  d-1\right)  \left(  d-4\right)  }\left(  -p\ell
^{2}\mathcal{R}^{\left(  0\right)  }+\frac{\left[  \left(  d-3\right)  \left(
d-4p\right)  +2\right]  }{2}tr\left[  \sigma^{\left(  2\right)  }\right]
 \right. \right. \nonumber\\
& \qquad\qquad\qquad\qquad\qquad\qquad \qquad\qquad + \left.  \left.  \frac{\left[  d-1-2p\left(  d-3\right)  \right]  }{2\left(
d-2\right)  }\ell^{2}\kappa_{a}^{\left(  \hat{\imath}\right)  a}\kappa
_{b}^{\left(  \hat{\imath}\right)  b}\right)  \right]  + \cdots \,,
\label{I_JM_expansion}%
\end{align}
after performing the radial integration up to the
cutoff scale $\rho=\epsilon$. 
In this expression, $C_{1}$ is the constant part, which for odd $d$ is
universal but for even $d$ depends on the choice of the cutoff $\epsilon$.

Consider next the Kounterterm contribution $I_{KT}$ in eq.\eqref{I_definitions}.
We show in Appendix \ref{AppendixF} that
\begin{align}
I_{KT}  &  =-\left(  \frac{d+1}{2}\right)  \left(  d-1\right)  c_{d}%
{\displaystyle\int\limits_{\partial\Sigma}}
d^{d-2}y\sqrt{\widetilde{\gamma}}PI\nonumber\\
&  =-%
{\displaystyle\int\limits_{\partial\Sigma}}
d^{d-2}y\frac{\sqrt{\sigma^{\left(  0\right)  }}}{\left(  d-2\right)
\epsilon^{\frac{d-2}{2}}}\sum\limits_{p=0}^{\frac{d-3}{2}}\frac{\left(
-1\right)  ^{p}\left(  p+1\right)  \left(  d-1\right)  !\alpha_{\left(
p+1\right)  }}{\left(  d-1-2p\right)  !\ell_{\text{eff}}^{2p-1}}%
\nonumber\\
& \qquad\qquad\qquad \times \left(  1-\epsilon\left[  \frac{\ell_{\text{eff}}^{2}}{2(d-4)}%
\mathcal{R}^{\left(  0\right)  }+\frac{1}{2}tr\left(  \sigma^{\left(
2\right)  }\right)  +\frac{\ell_{\text{eff}}^{2}}{2(d-2)}\kappa_{a}^{\left(
i\right)  a}\kappa_{b}^{\left(  i\right)  b}\right]  \right)  + \cdots \,,
\label{I_Kt}%
\end{align}
in both even and odd $d$.

Finally we  compute $S_{EE}^{\text{ren}}$ using the  expressions for $I_{JM}$ and $I_{KT}$ in eqs.\eqref{I_JM_expansion} and \eqref{I_Kt}. It is evident that the leading $O\left(  \epsilon
^{-\frac{d-2}{2}}\right)  $ divergences from $I_{JM}$ and $I_{KT}$ cancel each
other in full generality. 
 After some algebra, we obtain  
 \begin{align}
S_{EE}^{\text{ren}}  &  =\frac{I_{JM}+I_{KT}}{4G}\nonumber\\
&  =\frac{C_{1}}{4G}+\left(
{\displaystyle\sum\limits_{p=0}^{\left\lfloor \frac{d-2}{2}\right\rfloor }}
\frac{\left(  -1\right)  ^{p}\left(  p+1\right)  \left(  d-2\right)
!\alpha_{\left(  p+1\right)  }}{\left(  d-2-2p\right)  !\ell_{\text{eff}}%
^{2p}}\right)  S_{\text{diff}}+ \cdots \,,
\label{SEEDivCancel}
\end{align}
where%
\begin{equation}
S_{\text{diff}}=\frac{1}{4G}%
{\displaystyle\int\limits_{\partial\Sigma}}
\frac{d^{d-2}y\ell_{\text{eff}}\sqrt{\sigma^{\left(  0\right)  }}}{\left(
d-2\right)  \left(  d-4\right)  \epsilon^{\frac{d-4}{2}}}\left(  \frac{1}%
{2}\ell_{\text{eff}}^{2}\mathcal{R}^{\left(  0\right)  }+\left(  d-3\right)
tr\left[  \sigma^{\left(  2\right)  }\right]  +\frac{\left(  d-3\right)
\ell_{\text{eff}}^{2}}{2\left(  d-2\right)  }\kappa_{a}^{\left(  \hat{\imath
}\right)  a}\kappa_{b}^{\left(  \hat{\imath}\right)  b}\right)  \,,
\end{equation}
which can be rewritten as
\begin{equation}
S_{\text{diff}}=\frac{1}{4G}%
{\displaystyle\int\limits_{\partial\Sigma}}
\frac{d^{d-2}y\ell_{\text{eff}}^{3}\sqrt{\sigma^{\left(  0\right)  }}}{\left(
d-2\right)  \left(  d-4\right)  \epsilon^{\frac{d-4}{2}}} \left[  \delta
_{a}^{c}\delta_{b}^{d}\left(  W^{\left(  0\right)  }\right)_{\phantom{ab} cd}%
^{ab}-\left(  \kappa_{d}^{\left(  \hat{\imath}\right)  a}\kappa_{a}^{\left(
\hat{\imath}\right)  d}-\frac{\kappa_{d}^{\left(  \hat{\imath}\right)
d}\kappa_{a}^{\left(  \hat{\imath}\right)  a}}{\left(  d-2\right)  }\right)
\right]  \,,
\end{equation}
where $\left( W^{\left(  0\right)  }\right)_{\phantom{ab} cd}^{ab}$ is the Weyl tensor
computed from the metric $ \sigma^{\left(  0\right)  }_{ab}$ \cite{Anastasiou:2019ldc}.
$S_{\text{diff}}$ vanishes for AAdS manifolds with conformally flat boundaries
and for ball-shaped entangling regions, thus explicitly verifying the
cancellation of divergences up to next-to-leading order in this case.

\section{
Topological reinterpretation of HEE in odd-dimensional CFTs}
\label{TopologicalHEE}

Having obtained the renormalized HEE for CFTs dual to Lovelock-AdS gravity in
arbitrary dimension, we now focus on the odd-$d$ case. In this case, we
present a reinterpretation of the $S_{EE}^{\text{ren}}$ in terms of the sum of
a topological term, proportional to the Euler characteristic of $\Sigma$, and
a piece that is given by a polynomial of the AdS curvature of $\Sigma$,
analogous to the bulk $P\left(  \mathcal{F}\right)  $ in
eq.\eqref{P(F)BulkFixedJ}.

We start by considering $S_{EE}^{\text{ren}}$ and $S_{JM}\left[
\Sigma\right]  $ as given in eqs.\eqref{RenHEELovelock} and \eqref{ICod2Bulk},respectively.
Using  the Euler theorem \eqref{Euler-Theo}, we have
\begin{equation}%
{\displaystyle\int\limits_{\Sigma}}
E_{d-1}-%
{\displaystyle\int\limits_{\partial\Sigma}}
d^{d-2} y\,B_{d-2}=\left(  4\pi\right)  ^{\frac{d-1}{2}}\left(  \frac{d-1}{2}\right)
!\,\chi\left(  \Sigma\right)  \,,
\end{equation}
relating  the Euler density $E_{d-1}$ of the extremal surface $\Sigma$ and the Chern
form $B_{d-2}$ at its boundary. The topological number $\widehat{\tau}_{d}$  is defined as
\begin{equation}
\widehat{\tau}_{d}=-\frac{1}{4G}\frac{\left(  d+1\right)  }{2}c_{d}\left(
4\pi\right)  ^{\frac{d-1}{2}}\left(  \frac{d-1}{2}\right)  ! \,,
\label{TopologicalHat}%
\end{equation}
yielding
\begin{equation}
S\overset{\text{def.}}{=}4G\left(  S_{EE}^{\text{ren}}-\widehat{\tau}_{d}%
\chi\left(  \Sigma\right)  \right)  =%
{\displaystyle\int\limits_{\Sigma}}
{\displaystyle\sum\limits_{p=0}^{\frac{d-1}{2}}}
\alpha_{\left(  p+1\right)  }\left(  p+1\right)  L_{2p}\left[  \gamma\right] \,,
\end{equation}
upon exchanging $B_{d-2}$ in favor of $E_{d-1}$ in
eq.\eqref{RenHEELovelock}, and where we have renamed
\begin{align}
\alpha_{\frac{d+1}{2}}  &  =c_{d}=\frac{2}{d+1}%
{\displaystyle\sum\limits_{p=1}^{\frac{d-1}{2}}}
\frac{p\alpha_{p}\left(  -1\right)  ^{\frac{d+3}{2}-p}}{\left(  d+1-2p\right)
!}\left(  \ell_{\text{eff}}\right)  ^{\left(  d+1-2p\right)  } \,, \qquad 
L_{d-1}\left[  \gamma\right]     =E_{d-1}  \quad \,.
\end{align}
Recalling the eq.\eqref{I_definitions} and noting that 
\begin{equation}
\widehat{\mathcal{R}}_{\phantom{\alpha_{1}\alpha_{2}}\beta_{1}\beta_{2}%
}^{\alpha_{1}\alpha_{2}}=\mathcal{F}_{\beta_{1}\beta_{2}}^{\alpha_{1}\alpha_{2}}-\frac{1}%
{\ell_{\text{eff}}^{2}}\delta_{\beta_{1}\beta_{2}}^{\alpha_{1}\alpha_{2}},
\label{F_AdS_cod2}%
\end{equation}
relates the  Riemann tensor $\widehat{\mathcal{R}}_{\phantom{\alpha_{1}\alpha_{2}}\beta_{1}\beta_{2}%
}^{\alpha_{1}\alpha_{2}}$ to the  AdS curvature $\mathcal{F}_{\beta_{1}\beta_{2}%
}^{\alpha_{1}\alpha_{2}}$ on $\Sigma$, we can write
\begin{align}%
{\displaystyle\int\limits_{\Sigma}}
&d^{d-1}y\sqrt{\gamma}P_{\left(  d-1\right)  }\left(  \mathcal{F}\right)  \nonumber\\
 &  =%
{\displaystyle\int\limits_{\Sigma}}
d^{d-1}y\sqrt{\gamma}%
{\displaystyle\sum\limits_{p=0}^{\frac{d-1}{2}}}
\frac{\alpha_{\left(  p+1\right)  }\left(  p+1\right)  }{2^{p}} 
   \delta_{\beta_{1}\cdots\beta_{2p}}^{\alpha_{1}\cdots\alpha_{2p}}\left(
\mathcal{F}_{\alpha_{1}\alpha_{2}}^{\beta_{1}\beta_{2}}-\frac{2}%
{\ell_{\text{eff}}^{2}}\delta_{\alpha_{1}}^{\beta_{1}}\delta_{\alpha_{2}%
}^{\beta_{2}}\right)  \cdots\left(  \mathcal{F}_{\alpha_{2p-1}\alpha_{2p}%
}^{\beta_{2p-1}\beta_{2p}}-\frac{2}{\ell_{\text{eff}}^{2}}\delta
_{\alpha_{2p-1}}^{\beta_{2p-1}}\delta_{\alpha_{2p}}^{\beta_{2p}}\right)  \nonumber\\
&={\displaystyle\int\limits_{\Sigma}}
d^{d-1}y\sqrt{\gamma}%
{\displaystyle\sum\limits_{p=0}^{\frac{d-1}{2}}}
{\displaystyle\sum\limits_{j=0}^{p}}
\frac{p!\left(  d-1-2j\right)  !\left(  -1\right)  ^{p-j}\left(  p+1\right)
\alpha_{\left(  p+1\right)  }}{j!\left(  p-j\right)  !2^{j}\left(
d-1-2p\right)  !\ell_{\text{eff}}^{2\left(  p-j\right)  }}\delta_{\beta
_{1}\cdots\beta_{2j}}^{\alpha_{1}\cdots\alpha_{2j}}\mathcal{F}_{\alpha
_{1}\alpha_{2}}^{\beta_{1}\beta_{2}}\cdots\mathcal{F}_{\alpha_{2j-1}%
\alpha_{2j}}^{\beta_{2j-1}\beta_{2j}} \nonumber\\
&={\displaystyle\int\limits_{\Sigma}}
d^{d-1}y\sqrt{\gamma}%
{\displaystyle\sum\limits_{j=0}^{\frac{d-1}{2}}}
\widehat{c}_{j}\delta_{\beta_{1}\cdots\beta_{2j}}^{\alpha_{1}\cdots\alpha
_{2j}}\mathcal{F}_{\alpha_{1}\alpha_{2}}^{\beta_{1}\beta_{2}}\cdots
\mathcal{F}_{\alpha_{2j-1}\alpha_{2j}}^{\beta_{2j-1}\beta_{2j}} \,,
\end{align}
where the simplification proceeds in a manner analogous to that of  the bulk $P\left(  \mathcal{F}\right)  $ in section
\ref{P(F)_Bulk}, and where 
\begin{equation}
\widehat{c}_{j}=%
{\displaystyle\sum\limits_{p=j}^{\frac{d-1}{2}}}
\frac{p!\left(  d-1-2j\right)  !\left(  -1\right)  ^{p-j}\left(  p+1\right)
\alpha_{\left(  p+1\right)  }}{j!\left(  p-j\right)  !2^{j}\left(
d-1-2p\right)  !\ell_{\text{eff}}^{2\left(  p-j\right)  }} \,,
\end{equation}
is the coefficient of the $\mathcal{F}^{j}$ term.

Note that the co-dimension 2 $\widehat{c}_{j}$ coefficients of $S$ can be directly
related to the bulk coefficients $c_{j}$ of the bulk $P\left(  \mathcal{F}%
\right)  $ given in eq.\eqref{Cj}. In particular, we have%
\begin{equation}
\widehat{c}_{j}=2\left(  j+1\right)  c_{j+1}, \label{P(F)CoeffRelations}%
\end{equation}
and therefore
\begin{equation}%
{\displaystyle\int\limits_{\Sigma}}
d^{d-1}y\sqrt{\gamma}P_{\left(  d-1\right)  }\left(  \mathcal{F}\right)  =%
{\displaystyle\int\limits_{\Sigma}}
d^{d-1}y\sqrt{\gamma}%
{\displaystyle\sum\limits_{j=1}^{\frac{d-1}{2}}}
2\left(  j+1\right)  c_{j+1}\delta_{\beta_{1}\cdots\beta_{2j}}^{\alpha
_{1}\cdots\alpha_{2j}}\mathcal{F}_{\alpha_{1}\alpha_{2}}^{\beta_{1}\beta_{2}%
}\cdots\mathcal{F}_{\alpha_{2j-1}\alpha_{2j}}^{\beta_{2j-1}\beta_{2j}}
\label{Cod2P(F)}%
\end{equation}
since $\widehat{c}_{0} = 2 c_{1}=0$, as shown in  \eqref{c_1}.
Hence we finally obtain 
\begin{equation}
S_{EE}^{\text{ren}}=\frac{1}{4G}%
{\displaystyle\int\limits_{\Sigma}}
d^{d-1}y\sqrt{\gamma}P_{\left(  d-1\right)  }\left(  \mathcal{F}\right)
+\widehat{\tau}_{d}\chi\left(  \Sigma\right)  
\label{S_EE^renTopological}%
\end{equation}
for the renormalized HEE.

The reinterpretation of $S_{EE}^{\text{ren}}$ given in
eq.\eqref{S_EE^renTopological} extends the results previously obtained  for Einstein-AdS, in refs.
\cite{Anastasiou:2017xjr,Anastasiou:2018rla},
to the entire Lovelock class. Namely,
the rewriting of the universal part of the HEE in terms of a topological invariant and a curvature-dependent term is a general feature.

Also, the co-dimension 2 $P\left(  \mathcal{F}\right)  $ obtained in eq.\eqref{Cod2P(F)} has some noteworthy properties. For instance, comparing this object with the corresponding one coming from the bulk as given in
eq.\eqref{BulkP(F)}, we notice that $P\left(  \mathcal{F}\right)  $ has the
expected self-replicating property when going to co-dimension 2, in analogy
with the Einstein-AdS case \cite{Anastasiou:2018mfk}. However, one not
only has to replace $d$ by $\left(  d-2\right)  $ but also $\alpha_{p}$ by
$\left(  p+1\right)  \alpha_{p+1}$, which comes from the fact that the JM
functional has the form of a derivarive with respect to the Riemann curvature
of the Lovelock Lagrangian, but evaluated intrinsically on the extremal surface.

For the degenerate cases it is trivial to relate the
$\widehat{c}_{j}$ to the degeneracy conditions using eq.\eqref{P(F)CoeffRelations}.
In particular, from eqs.\eqref{c_2} and \eqref{c_3} in Appendix \ref{AppendixC}, we have
\begin{align}
\widehat{c}_{1}  &  =4\,c_{2}=\frac{\ell_{\text{eff}}^{2}}{2^{2}\left(
d-2\right)  }\Delta^{\left(  1\right)  }\left(  \ell_{\text{eff}}^{-2}\right)
,\nonumber\\
\widehat{c}_{2}  &  =6\,c_{3}=-\frac{\ell_{\text{eff}}^{2}}{\left(  d-2\right)
\left(  d-4\right)  2^{3}}\left(  \frac{1}{\left(  d-3\right)  }%
\Delta^{\left(  2\right)  }+\frac{\ell_{\text{eff}}^{2}}{4}\Delta^{\left(
1\right)  }\right)   \,,
\end{align}
and in the general case
\begin{equation}
\widehat{c}_{j}=2\left(  j+1\right)  c_{j+1}=2\left(  j+1\right)
{\displaystyle\sum\limits_{i=1}^{j}}
p_{\left(  j+1,i\right)  }\,\Delta^{\left(  i\right)  },
\end{equation}
from eq.\eqref{GeneralDegeneracy} for some coefficients $p_{\left(  j+1,i\right)  }$ (given in eq.\eqref{P_ij}). 
In a $k-$fold degenerate theory, the lowest order in
$\mathcal{F}$ of the co-dimension 2 polynomial $P_{\left(  d-1\right)  }\left(  \mathcal{F}%
\right)  $ will be $\mathcal{F}^{k+1}$, in an analogous manner to the Noether
prepotential discussed in Appendix \ref{AppendixD}. Therefore we can write
\begin{equation}
P_{\left(  d-1\right)  }\left(  \mathcal{F}\right)  =%
{\displaystyle\sum\limits_{j=1}^{\frac{d-1}{2}}}
{\displaystyle\sum\limits_{i=1}^{j}}
2\left(  j+1\right)  \left(  p_{\left(  j+1,i\right)  }\Delta^{\left(
i\right)  }\right)  \delta_{\beta_{1}\cdots\beta_{2j}}^{\alpha_{1}\cdots
\alpha_{2j}}\mathcal{F}_{\alpha_{1}\alpha_{2}}^{\beta_{1}\beta_{2}}%
\cdots\mathcal{F}_{\alpha_{2j-1}\alpha_{2j}}^{\beta_{2j-1}\beta_{2j}} \,,
\end{equation}
in terms of
the degeneracy conditions $\Delta^{\left(  k\right)  }$.

Note that in a $k$-degenerate theory, the co-dimension 2 polynomial $P_{\left(  d-1\right)
}\left(  \mathcal{F}\right)  $ can be written as%
\begin{equation}
P_{\left(  d-1\right)  }\left(  \mathcal{F}\right)  =\widehat{c}_{k+1}%
\delta_{\beta_{1}\cdots\beta_{2k+2}}^{\alpha_{1}\cdots\alpha_{2k+2}%
}\mathcal{F}_{\alpha_{1}\alpha_{2}}^{\beta_{1}\beta_{2}}\cdots\mathcal{F}%
_{\alpha_{2k+1}\alpha_{2k+2}}^{\beta_{2k+1}\beta_{2k+2}}+%
{\displaystyle\sum\limits_{j=k+2}^{\frac{d-1}{2}}}
\widehat{c}_{j}\delta_{\beta_{1}\cdots\beta_{2j}}^{\alpha_{1}\cdots\alpha
_{2j}}\mathcal{F}_{\alpha_{1}\alpha_{2}}^{\beta_{1}\beta_{2}}\cdots
\mathcal{F}_{\alpha_{2j-1}\alpha_{2j}}^{\beta_{2j-1}\beta_{2j}},
\label{HEEP(F)Deg}
\end{equation}
where%
\begin{equation}
\widehat{c}_{k+1}=2\left(  k+2\right)  p_{\left(  k+2,k+1\right)  }\,%
\Delta^{\left(  k+1\right)  },
\end{equation}
for $p_{\left(  k+2,k+1\right)  }$ as defined in eq.(\ref{DegP}).

We close this section by commenting that   $S_{EE}^{\text{ren}}$ for degenerate theories
exhibits an interesting robustness property under shape deformations of the
entangling region. What occurs is that, as for the
Einstein-AdS case (in $d=3$) \cite{Anastasiou:2020smm}, the topological part of the renormalized HEE is
unchanged under such deformations, whereas the $P_{\left(  d-1\right)
}\left(  \mathcal{F}\right)  $ part changes  as
$\varepsilon^{2}$ to   lowest order in the   deformation parameter $\varepsilon$.   This is a result of the fact $\mathcal{F}$ is second order in $\varepsilon$ and that  $P_{\left(  d-1\right)  }\left(
\mathcal{F}\right)  $ is of order $\mathcal{F}^{1}$.  However, for a $k-$fold degenerate
Lovelock theory,   $P_{\left(  d-1\right)  }\left(
\mathcal{F}\right)  $ is of order
$\mathcal{F}^{k+1}$ and higher, and therefore $P_{\left(
d-1\right)  }\left(  \mathcal{F}\right)  $  changes as $\varepsilon
^{2k+2}$ to the leading order in $\varepsilon$. 

This means that the
renormalized EE of a CFT dual to a degenerate theory is robust under shape
deformations, and increasingly so  the higher its degeneracy. It is
interesting to conjecture that this robustness of $S_{EE}^{\text{ren}}$ under
such deformations for degenerate Lovelock theories could constrain
higher-order correlators in the dual CFT. However, the
study of shape deformations falls outside the scope of this work and will be
pursued in a future paper.

\section{Renormalized HEE for ball-shaped regions and
$C-$function candidates}
\label{BallHEE}

We now proceed to compute the renormalized HEE for ball-shaped entangling
surfaces. This case is important as the universal part of the HEE, which is
equal to the renormalized value, is directly related to the $C%
$-function candidate of the CFT \cite{Anastasiou:2019ldc}. The extremal surface $\Sigma$ of the JM
functional in this geometry is the hemisphere in the bulk, as discussed in
Appendix \ref{AppendixE}. We analyze the odd-$d$ and even-$d$ cases
separately, as the expressions for $S_{EE}^{\text{ren}}$ are different.

We start with odd-dimensional CFTs. In this case $\Sigma$ (as defined in  eq.(\ref{AnsatzSigma}) in Appendix \ref{AppendixE}) is a constant-curvature surface, topologically equivalent to a ball. It therefore has vanishing AdS
curvature $\mathcal{F}_{AdS}$ and Euler characteristic $\chi\left(\Sigma\right)  =1$. Using
eq.(\ref{S_EE^renTopological}) we directly obtain
\begin{align}
S_{EE}^{\text{ren}}  &  =\widehat{\tau}_{d}\nonumber\\
&  =%
{\displaystyle\sum\limits_{p=1}^{\frac{d-1}{2}}}
\frac{\left(  -1\right)  ^{\frac{d+1}{2}-p}\left(  4\pi\right)  ^{\frac
{d-1}{2}}\left(  \frac{d-1}{2}\right)  !p\alpha_{p}}{4G_{N}\left(
d+1-2p\right)  !}\left(  \ell_{\text{eff}}\right)  ^{\left(  d+1-2p\right)
}\nonumber\\
&  =\left(  -1\right)  ^{\frac{d-1}{2}}\left(
{\displaystyle\sum\limits_{p=0}^{\frac{d-3}{2}}}
\frac{\left(  p+1\right)  \alpha_{p+1}\left(  d-1\right)  !\left(  -1\right)
^{p}}{\left(  d-1-2p\right)  !\ell_{\text{eff}}^{2p}}\right)  \left(
\frac{\left(  4\pi\right)  ^{\frac{d-1}{2}}\left(  \frac{d-1}{2}\right)
!}{4G_{N}\left(  d-1\right)  !}\ell_{\text{eff}}^{d-1}\right)  ,
\label{SRenBallOdd-d}%
\end{align}
using the definitions of $\widehat{\tau}_{d}$ and $c_{d}$ given in eqs.\eqref{TopologicalHat} and \eqref{KountertermCoupling}. This factorized form is in accord with   previous
results for the Einstein-AdS case \cite{Anastasiou:2017xjr,Anastasiou:2018rla}. Indeed, for Einstein gravity, the first factor in the last expression is identically equal to
one. This first prefactor corresponds to the usual constant encountered in the linearization of Lovelock gravity.

Note that the result of eq.(\ref{SRenBallOdd-d}) implies that the coefficient
$\widehat{\tau}_{d}$ of the topological term in eq.(\ref{S_EE^renTopological})
can be written as%
\begin{equation}
\widehat{\tau}_{d}=\left(  -1\right)^{\frac{d-1}{2}}F_{L} =  \left(  -1\right)^{\frac{d-1}{2}} Q_{L}F_{EH} \,,
\end{equation} 
where%
\begin{align} 
F_{EH}  &  =\left(  \frac{\left(  4\pi\right)  ^{\frac{d-1}{2}}\left(
\frac{d-1}{2}\right)  !}{4G_{N}\left(  d-1\right)  !}\ell_{\text{eff}}%
^{d-1}\right)  \,, \qquad
Q_{L}    =\left(
{\displaystyle\sum\limits_{p=0}^{\left\lfloor \frac{d-2}{2}\right\rfloor }}
\frac{\left(  p+1\right)  \alpha_{p+1}\left(  d-1\right)  !\left(  -1\right)
^{p}}{\left(  d-1-2p\right)  !\ell_{\text{eff}}^{2p}}\right)  , \label{F_L}%
\end{align}
so that $F_{EH}$ is the usual generalized $F-$quantity ($C-$function
candidate) for odd dimensional CFTs dual to Einstein-AdS gravity. The quantity $Q_{L}$
is the prefactor that depends on the Lovelock couplings. Thus, it is natural to
conjecture that the $C-$function candidate for odd-$d$ CFTs dual to
Lovelock-AdS gravity is given directly by $F_{L}$.

For even-$d$ CFTs, dual to odd-$D$ Lovelock-AdS gravity, the universal part of
the HEE corresponds to the coefficient of the logarithmically divergent part,
which is proportional to the type-A anomaly coefficient of the CFT. To compute
the universal part we start from the expression \eqref{RenHEELovelock} for $S_{EE}^{\text{ren}}$, and note
that for a spherical hemisphere in the bulk (the 
 $\Sigma$ of eq.\eqref{AnsatzSigma}) its induced metric \eqref{IntrinsicMetricHemisphere} satisfies
\begin{equation}
\widehat{\mathcal{R}}_{\phantom{\alpha_{1}\alpha_{2}}\beta_{1}\beta_{2}}^{\alpha_{1}\alpha_{2}}=-\frac
{1}{\ell_{\text{eff}}^{2}}\delta_{\beta_{1}\beta_{2}}^{\alpha_{1}\alpha_{2}} \,,
\end{equation}
since it is a
constant curvature manifold. Using the definition of the $L_{2p}$ Lovelock densities in eq.\eqref{Lovelock_Density} and the delta identities of Appendix \ref{AppendixA},
we obtain
\begin{align}
L_{2p}\left[  \gamma\right]   &  =\frac{1}{2^{p}}\left(  -\frac{2}%
{\ell_{\text{eff}}^{2}}\right)  ^{p}\delta_{\beta_{1}\cdots\beta_{2p}}%
^{\alpha_{1}\cdots\alpha_{2p}}\delta_{\alpha_{1}}^{\beta_{1}}\cdots
\delta_{\alpha_{2p}}^{\beta_{2p}}\sqrt{\gamma}  =\frac{\left(  d-1\right)  !}{\left(  d-1-2p\right)  !}\frac{\left(
-1\right)  ^{p}}{\ell_{\text{eff}}^{2p}}\sqrt{\gamma}.
\end{align}
Therefore, the JM functional becomes%
\begin{align}
S_{JM}  &  =\frac{1}{4G}%
{\displaystyle\int\limits_{\Sigma}}
\sqrt{\gamma}d^{d-1}y%
{\displaystyle\sum\limits_{p=0}^{\left(  d-2\right)  /2}}
\alpha_{\left(  p+1\right)  }\left(  p+1\right)  \frac{\left(  d-1\right)
!}{\left(  d-1-2p\right)  !}\frac{\left(  -1\right)  ^{p}}{\ell_{\text{eff}%
}^{2p}}\nonumber\\
&  =\left(
{\displaystyle\sum\limits_{p=0}^{\left(  d-2\right)  /2}}
\alpha_{\left(  p+1\right)  }\left(  p+1\right)  \frac{\left(  d-1\right)
!}{\left(  d-1-2p\right)  !}\frac{\left(  -1\right)  ^{p}}{\ell_{\text{eff}%
}^{2p}}\right)  \frac{\text{Area}\left(  \Sigma\right)  }{4G}.
\label{S_JM-Vol}%
\end{align}
On the other hand, the contribution to the entropy from the Kounterterms is given by
\begin{equation}
S_{EE}^{\text{KT}}=\frac{d}{2}\frac{c_{d}}{4G}%
{\displaystyle\int\limits_{\partial\Sigma}}
d^{d-2} y\,B_{d-2} \,,
\end{equation}
where $c_{d}$ is defined in  eq.\eqref{KountertermCoupling}. The universal
part of the HEE is thus given by%
\begin{align}
S_{EE}^{\text{Univ}}  &  =S_{JM}+S_{EE}^{\text{KT}}\nonumber\\
&  = \left( \left(
{\displaystyle\sum\limits_{p=0}^{\left(  d-2\right)  /2}}
\alpha_{\left(  p+1\right)  }\left(  p+1\right)  \frac{\left(  d-1\right)
!}{\left(  d-1-2p\right)  !}\frac{\left(  -1\right)  ^{p}}{\ell_{\text{eff}%
}^{2p}}\right)  \frac{\text{Area}\left(  \Sigma\right)  }{4G} \right.
\nonumber\\
& \qquad \qquad +  \left.  \frac{1}{4G}\left[  \frac{\left(  d-1\right)  !}{2^{d-2}\left[
\left(  \frac{d}{2}-1\right)  !\right]  ^{2}}\right]
{\displaystyle\sum\limits_{p=0}^{\frac{d-2}{2}}}
\frac{\left(  p+1\right)  \alpha_{p+1}\left(  -1\right)  ^{\frac{d}{2}-p}%
}{\left(  d-1-2p\right)  !}\left(  \ell_{\text{eff}}\right)  ^{\left(
d-2p-2\right)  }%
{\displaystyle\int\limits_{\partial\Sigma}}
B_{d-2} \right)  \nonumber\\
&  =\frac{1}{4G}\left(
{\displaystyle\sum\limits_{p=0}^{\frac{d-2}{2}}}
\frac{\left(  p+1\right)  \alpha_{p+1}\left(  d-1\right)  !\left(  -1\right)
^{p}}{\left(  d-1-2p\right)  !\ell_{\text{eff}}^{2p}}\right)  \left(
\text{Area}\left(  \Sigma\right)  -\frac{d}{2}\left[  \frac{\left(  -1\right)
^{\frac{d}{2}}\left(  \ell_{\text{eff}}\right)  ^{\left(  d-2\right)  }%
}{2^{d-3}d\left[  \left(  \frac{d}{2}-1\right)  !\right]  ^{2}}\right]
{\displaystyle\int\limits_{\partial\Sigma}}
B_{d-2}\right) \nonumber\\
&=\left(
{\displaystyle\sum\limits_{p=0}^{\frac{d-2}{2}}}
\frac{\left(  p+1\right)  \alpha_{p+1}\left(  d-1\right)  !\left(  -1\right)
^{p}}{\left(  d-1-2p\right)  !\ell_{\text{eff}}^{2p}}\right)  \frac
{\text{Area}_{\text{Univ}}\left(  \Sigma\right)  }{4G}
\label{HEELball}
\end{align}
where
\begin{equation}
\text{Area}_{\text{Univ}}\left(  \Sigma\right)  \overset{\text{def.}}%
{=}\text{Area}\left(  \Sigma\right)  -\frac{d}{2}\left[  \frac{\left(
-1\right)  ^{\frac{d}{2}}\left(  \ell_{\text{eff}}\right)  ^{\left(
d-2\right)  }}{2^{d-3}d\left[  \left(  \frac{d}{2}-1\right)  !\right]  ^{2}%
}\right]
{\displaystyle\int\limits_{\partial\Sigma}}
d^{d-2}y\,B_{d-2}  
\label{AreaRen}%
\end{equation}
is independent of any Lovelock factors.

In the Einstein-AdS case the
universal part of the HEE is precisely equal to $\frac{\text{Area}_{\text{Univ}}\left(  \Sigma\right)  }{4G}$, such that Area$_{\text{Univ}}$
exactly matches the universal (logarithmically divergent) part of the area for
the minimal (RT) surface \cite{Anastasiou:2019ldc}. In other words
\begin{equation}
\frac{\text{Area}_{\text{Univ}}\left(  \Sigma\right)  }{4G}=\left(  -1\right)
^{\frac{d}{2}+1}4\left[  \frac{\ell_{\text{eff}}^{d-1}\pi^{\frac{d}{2}-1}%
}{8G\left(  \frac{d}{2}-1\right)  !}\right]  \ln\left(  \frac{L}{\delta
}\right)  ,
\end{equation}
where $\delta$ is a cutoff scale in the Poincaré coordinate $z\overset
{\text{def.}}{=}\ell\sqrt{\rho}$, which scales as a length and $L$ is the
radius of the entangling region in the CFT.  We see from eq.\eqref{HEELball} that this quantity likewise
governs the universal part of the HEE in Lovelock gravity, so that
\begin{equation}
S_{EE}^{\text{Univ}}=\left(  -1\right)  ^{\frac{d}{2}+1}4\left(
{\displaystyle\sum\limits_{p=0}^{\frac{d-2}{2}}}
\frac{\left(  p+1\right)  \alpha_{p+1}\left(  d-1\right)  !\left(  -1\right)
^{p}}{\left(  d-1-2p\right)  !\ell_{\text{eff}}^{2p}}\right)  \left[
\frac{\ell_{\text{eff}}^{d-1}\pi^{\frac{d}{2}-1}}{8G\left(  \frac{d}%
{2}-1\right)  !}\right]  \ln\left(  \frac{L}{\delta}\right)  \,,
\label{SReneven-dimensional}
\end{equation}
for ball-shaped entangling regions in even-$d$ CFTs.
We recognize that the factor with the sum is again the usual factor $Q_{L}$
(as given in eq.(\ref{F_L})) that appears in solutions to Lovelock gravity. This is identically equal to
one in the case of Einstein-AdS gravity, recovering   previous results in the Einstein-AdS case \cite{Anastasiou:2019ldc}.

Note that eq.(\ref{SReneven-dimensional}) can be rewritten as%
\begin{equation}
S_{EE}^{\text{Univ}}=\left(  -1\right)  ^{\frac{d}{2}+1}4A_{L}\ln\left(
\frac{L}{\delta}\right)  ,
\end{equation}
such that%
\begin{align}
A_{L}  &  =Q_{L}A_{EH} \,, \qquad A_{EH}    =\frac{\ell_{\text{eff}}^{d-1}\pi^{\frac{d}{2}-1}}{8G\left(
\frac{d}{2}-1\right)  !}, 
\label{A_L}%
\end{align}
with   $A_{EH}$ being the type-A anomaly
coefficient ($C-$ function candidate) of even-$d$ CFTs dual to Einstein-AdS.
Therefore, it is natural to conjecture that for even-$d$ CFTs dual to
Lovelock-AdS, the type-A anomaly coefficient is given directly by $A_{L}$.\footnote{The $C$-function candidate for generic $\mathcal{L}$(Riemann) theories has been previously written, in ref.\cite{Myers:2010tj}, as proportional to the Lagrangian evaluated on the AdS vacuum with effective radius $\ell_{\text{eff}}$. It can be shown that our expression matches said results.}

\section{Conclusions}\label{Conclusions}

We have shown how  the renormalization of HEE for CFTs dual to Lovelock gravity can be carried out using the Kounterterm procedure. The computation takes advantage of the self-replicating property of Lovelock densities $L_{2p}$ \cite{Kastikainen:2020auf} and the one of the extrinsic counterterms $B_{d}$ \cite{Anastasiou:2019ldc}, when evaluated on squashed cones. In particular, when evaluated on a cone, both  $L_{2p}$ and the $B_{d}$ split into a regular contribution and the corresponding co-dimension 2 term localized at the conical singularity. This allows us to cast the renormalized HEE  into the form \eqref{RenHEELovelock},   in agreement with the result of ref.\cite{Hung:2011xb} for the bulk term of the JM entropy functional. Furthermore, our result shows that the co-dimension 2 Kounterterm added at the boundary $\partial \Sigma$ of the extremal surface is the  structure that renormalizes the entanglement entropy.

In studying the renormalization of the gravity theory, we were able to write the Lovelock Lagrangian ${\cal L}$ for even-dimensional bulk manifolds as a polynomial of the AdS curvature $\mathcal{F}$ \eqref{BulkP(F)}. This form is convenient as the coefficients of the different powers of $\mathcal{F}$ are linear combinations of the degeneracy conditions \eqref{GenericCoeff}, extending the analysis of ref.\cite{Arenas-Henriquez:2019rph} for the Noether prepotential used in the computation of asymptotic charges. Indeed,    writing  the Lagrangian   in terms of the degeneracy conditions makes it clear that any variation of ${\cal L}$  and, therefore, the Noether prepotential, will inherit the same property. 
Furthermore, the  relation between  the degeneracy conditions  and the coefficients of powers of $\mathcal{F}$ is also valid for the renormalized JM functional for odd-dimensional CFTs dual to Lovelock gravity. 

It is evident from  \eqref{S_EE^renTopological} that the finite entanglement entropy contains a topological contribution proportional to the Euler characteristic of the extremal surface. The proportionality constant corresponds to the generalized F-quantity of the theory. Furthermore, a purely geometric term arises at   finite order, which can be expressed as a polynomial in  $\mathcal{F}$. This extends the known relation between the finite part of the HEE for spherical entangling regions in Einstein gravity and the F-quantity to different shapes and higher Lovelock densities. 

For generic Lovelock gravity dual to CFTs of both even and odd dimensions, we have shown the finiteness of our renormalized HEE functional \eqref{SEEDivCancel} . Our analysis applies to Lovelock densities of arbitrary  degree in the curvature, and the cancellation of the leading-order term was shown in full generality. The cancellation of the next-to-leading order divergence was verified in the case of manifolds with conformally flat boundaries and for spherical entangling regions. 

Our results demonstrate that the Kounterterm scheme efficiently isolates the universal part of the HEE, for both even and odd-dimensional cases. This universal part, for spherical entangling regions, corresponds to the C-function candidate. For odd-dimensional CFTs, this candidate is the F-quantity (proportional to the CFT partition function evaluated on a sphere), whereas for even-dimensional CFTs it is the type-A anomaly coefficient. We obtained explicit formulas for both quantities in eqs. \eqref{F_L} and \eqref{A_L}, respectively, where we note  that the C-function candidate is always proportional to the one for Einstein-AdS gravity, with a coupling-dependent factor,  commensurate with the recent literature \cite{Bueno:2020uxs,Anastasiou:2021swo}. 

We emphasize that Kounterterm method  constitutes a notable improvement over other approaches as it is non-perturbative, neither assuming that the Lovelock couplings are small, nor that the Lovelock theory has an Einstein behavior. An example of the latter is given by degenerate Lovelock theories, where the method is still applicable for isolating the universal part of the HEE. It is interesting to note (as mentioned in eq.\eqref{HEEP(F)Deg}) that the resulting renormalized JM functional for odd-dimensional CFTs  is more robust under deformations of the entangling region in the degenerate cases. The reason for this is that the leading power (in $\mathcal{F}$) of the polynomial form of the JM functional depends on the degeneracy condition; for a $k$-degenerate theory it is $\mathcal{F}^{k+1}$.  Since  the topological part is not affected by continuous deformations of the entangling surface, the resulting effect will enter only through the change in $\mathcal{F}$, which to   leading order  is quadratic in the deformation parameter. It would be interesting to explore   the consequences of this feature, but an analysis of shape deformations falls outside the scope of the present paper.

\acknowledgments

We thank Alberto G\"uijosa for interesting discussions and feedback. The work of GA and RO was funded in part by FONDECYT grants No. 3190314 \textit{Holographic Complexity from Anti-de Sitter gravity} and No. 1170765 \textit{Boundary dynamics in anti-de Sitter gravity and gauge/gravity duality}. The work of IJA is funded by ANID, REC Convocatoria Nacional Subvenci\'on a Instalaci\'on en la Academia Convocatoria A\~no 2020, Folio PAI77200097. The work of RBM was supported in part by the Natural Sciences and Engineering Research Council of Canada.

\appendix
\section{Generalized Kronecker delta identities}\label{AppendixA}

In the main text, we use three identities involving the generalized Kronecker
delta. The most important one is%
\begin{equation}
\delta_{\nu_{1}\cdots\nu_{k}}^{\mu_{1}\cdots\mu_{k}}\delta_{\mu_{k-p+1}}%
^{\nu_{k-p+1}}\cdots\delta_{\mu_{k}}^{\nu_{k}}=\frac{\left(  r-k+p\right)
!}{\left(  r-k\right)  !}\delta_{\nu_{1}\cdots\nu_{k-p}}^{\mu_{1}\cdots
\mu_{k-p}}, \label{ContractDeltaIdentity}%
\end{equation}
which allows to lower its rank by contracting it with a sequence of
rank-1 Kronecker deltas. In the expression
above, $r$ is the range of the indices, $k$ is the rank of the generalized Kronecker delta,  and $p$ is the number of rank-1 deltas present in the sequence.

The second identity is%
\begin{equation}
\delta_{\nu_{1}\cdots\nu_{k}}^{\mu_{1}\cdots\mu_{k}}\delta_{\mu_{1}\mu_{2}%
}^{\nu_{1}\nu_{2}}\cdots\delta_{\mu_{2q-1}\mu_{2q}}^{\nu_{2q-1}\nu_{2q}}%
=2^{q}\delta_{\nu_{1}\cdots\nu_{k}}^{\mu_{1}\cdots\mu_{k}}\delta_{\mu_{1}%
}^{\nu_{1}}\cdots\delta_{\mu_{2q}}^{\nu_{2q}}, \label{DissolveDeltaIdentity}%
\end{equation}
which is simply a consequence of the antisymmetry of the rank-2 delta and of
the overall contaction implemented by the rank-$k$ delta.

The third identity follows directly from the previous two identities. It
states that%
\begin{gather}
N=\left\lfloor \frac{d}{2}\right\rfloor \text{ },\text{ }r=d+1,\nonumber\\
\delta_{\nu\nu_{1}\cdots\nu_{2p}}^{\mu\mu_{1}\cdots\mu_{2p}}=\frac{1}%
{2^{N-p}\left(  d-2p\right)  !}\delta_{\nu\nu_{1}\cdots\nu_{2N}}^{\mu\mu
_{1}\cdots\mu_{2N}}\delta_{\mu_{2p+1}\mu_{2p+2}}^{\nu_{2p+1}\nu_{2p+2}}%
\cdots\delta_{\mu_{2N-1}\mu_{2N}}^{\nu_{2N-1}\nu_{2N}}.
\label{BinomialDeltaIdentity}%
\end{gather}
In the previous expression, $\left\lfloor x\right\rfloor $ is the integer
floor of $x$ and $d+1$ is the range of the indices of the deltas. The proof of this identity has to be
done separately for odd and even $d$, but it is straightforward.

\section{Factorization of the Lovelock equation of motion}\label{AppendixB}

We begin by noting that the Lovelock EOM can be written as%
\begin{equation}
E_{\mu}^{\nu}=%
{\displaystyle\sum\limits_{p=0}^{N}}
\frac{\alpha_{p}}{2^{p+1}}\frac{1}{2^{N-p}\left(  d-2p\right)  !}\delta
_{\mu\mu_{1}\cdots\mu_{2N}}^{\nu\nu_{1}\cdots\nu_{2N}}\delta_{\nu_{2p+1}%
\nu_{2p+2}}^{\mu_{2p+1}\mu_{2p+2}}\cdots\delta_{\nu_{2N-1}\nu_{2N}}%
^{\mu_{2N-1}\mu_{2N}}R_{\phantom{\mu_{1}\mu_{2}}\nu_{1}\nu_{2}}^{\mu_{1}\mu_{2}}\cdots 
R_{\phantom{\mu_{2p-1}\mu_{2p}}\nu_{2p-1}\nu_{2p}}^{\mu_{2p-1}\mu_{2p}}=0 \,.
\label{B1}
\end{equation}
We have rewritten
$\delta_{\mu\mu_{1}\cdots\mu_{2p}}^{\nu\nu_{1}\cdots\nu_{2p}}$ of
eq.(\ref{LovelockEOM}) using the delta identity of
eq.(\ref{BinomialDeltaIdentity}). We now define%
\begin{equation}
A_{\nu_{1}\cdots\nu_{2N}}^{\mu_{1}\cdots\mu_{2N}}=%
{\displaystyle\sum\limits_{p=0}^{N}}
\frac{\alpha_{p}d!}{\left(  d-2p\right)  !} R_{\phantom{\mu_{1}\mu_{2}}\nu_{1}\nu_{2}}^{\mu_{1}\mu_{2}}\cdots 
R_{\phantom{\mu_{2p-1}\mu_{2p}}\nu_{2p-1}\nu_{2p}}^{\mu_{2p-1}\mu_{2p}}
 \delta_{\nu_{2p+1}%
\nu_{2p+2}}^{\mu_{2p+1}\mu_{2p+2}}\cdots\delta_{\nu_{2N-1}\nu_{2N}}%
^{\mu_{2N-1}\mu_{2N}}\,,%
\end{equation}
and thus, the EOM can be rewritten as%
\begin{equation}
\delta_{\mu\mu_{1}\cdots\mu_{2N}}^{\nu\nu_{1}\cdots\nu_{2N}}\frac{1}%
{d!2^{N+1}}A_{\nu_{1}\cdots\nu_{2N}}^{\mu_{1}\cdots\mu_{2N}}=0\,.
\end{equation}
The tensor $A$ can then be factorized in a straightforward manner 
\begin{align}
A_{\nu_{1}\cdots\nu_{2N}}^{\mu_{1}\cdots\mu_{2N}}  &  =%
{\displaystyle\sum\limits_{p=0}^{N}}
\frac{\alpha_{p}d!}{\left(  d-2p\right)  !}x_{\nu_{1}\nu_{2}}^{\mu_{1}\mu_{2}%
}\cdots x_{\nu_{2p-1}\nu_{2p}}^{\mu_{2p-1}\mu_{2p}}y_{\nu_{2p+1}\nu_{2p+2}%
}^{\mu_{2p+1}\mu_{2p+2}}\cdots y_{\nu_{2N-1}\nu_{2N}}^{\mu_{2N-1}\mu_{2N}} \,,
\end{align}
by writing
$$
  x_{\nu_{1}\nu_{2}}^{\mu_{1}\mu_{2}}\overset{\text{def.}}{=}
R_{\phantom{\mu_{1}\mu_{2}}\nu_{1}\nu_{2}}^{\mu_{1}\mu_{2}} \,, \qquad 
\text{ }y_{\nu_{1}\nu_{2}}^{\mu_{1}\mu_{2}%
}\overset{\text{def.}}{=}\delta_{\nu_{1}\nu_{2}}^{\mu_{1}\mu_{2}}  \quad .
$$
Then, $A$ , as a polynomial, can be decomposed into its roots as
\begin{align}
A_{\nu_{1}\cdots\nu_{2N}}^{\mu_{1}\cdots\mu_{2N}}  &  =\alpha_{N}d!\left(
x_{\nu_{1}\nu_{2}}^{\mu_{1}\mu_{2}}-\left(  x_{\left(  1\right)  }\right)
_{\nu_{1}\nu_{2}}^{\mu_{1}\mu_{2}}\right)  \cdots\left(  x_{\nu_{2N-1}\nu
_{2N}}^{\mu_{2N-1}\mu_{2N}}-\left(  x_{\left(  N\right)  }\right)
_{\nu_{2N-1}\nu_{2N}}^{\mu_{2N-1}\mu_{2N}}\right)  ,\nonumber\\
\left(  x_{\left(  i\right)  }\right)  _{\nu_{2i-1}\nu_{2i}}^{\mu_{2i-1}%
\mu_{2i}}  &  =-\lambda_{\left(  i\right)  }y_{\nu_{2i-1}\nu_{2i}}^{\mu
_{2i-1}\mu_{2i}},
\end{align}
where $\left\{  \lambda_{\left(  i\right)  }\right\}  $ are the solutions of%
\begin{equation}%
{\displaystyle\sum\limits_{p=0}^{N}}
\frac{\alpha_{p}d!}{\left(  d-2p\right)  !}\left(  -1\right)  ^{p}\lambda
^{p}=0.
\end{equation}
As a consequence, the equations of motion \eqref{B1} can be cast as
\begin{equation}
E_{\mu}^{\nu}=\delta_{\mu\mu_{1}\cdots\mu_{2N}}^{\nu\nu_{1}\cdots\nu_{2N}%
}\frac{\alpha_{N}}{ 2^{N+1}} \left(  R_{\phantom{\mu_{1}\mu_{2}}\nu_{1}\nu_{2}}^{\mu_{1}\mu_{2}%
}+\lambda_{\left(  1\right)  }\delta_{\nu_{1}\nu_{2}}^{\mu_{1}\mu_{2}}\right)
\cdots\left(  R_{\phantom{\mu_{2N-1}\mu_{2N}}\nu_{2N-1}\nu_{2N}}^{\mu_{2N-1}\mu_{2N}}+\lambda_{\left(
N\right)  }\delta_{\nu_{2N-1}\nu_{2N}}^{\mu_{2N-1}\mu_{2N}}\right)  =0
\label{B7}
\end{equation}
where $\lambda_{\left(  i\right)  }$ is solution of%
\begin{equation}%
{\displaystyle\sum\limits_{p=0}^{N}}
\frac{\alpha_{p}d!}{\left(  d-2p\right)  !}\left(  -1\right)  ^{p}\lambda
^{p}=-d\left(  d-1\right)  \Delta\left(  \lambda\right)  =0,
\end{equation}
and the polynomial $ \Delta\left(  \lambda\right)  $ is defined in eq.(\ref{CharPoly}).

Hence, the EOM \eqref{B7} in factorized form is  
\begin{equation}
E_{\mu}^{\nu}=\frac{\alpha_{N}}{2^{N+1}}\delta_{\mu\mu_{1}\cdots\mu_{2N}}%
^{\nu\nu_{1}\cdots\nu_{2N}}\left(  R_{\phantom{\mu_{1}\mu_{2}}\nu_{1}\nu_{2}}^{\mu_{1}\mu_{2}}%
+\frac{1}{\ell_{\text{eff}(1)}^{2}}\delta_{\nu_{1}\nu_{2}}^{\mu_{1}\mu_{2}%
}\right)  \cdots\left(  R_{\phantom{\mu_{2N-1}\mu_{2N}}\nu_{2N-1}\nu_{2N}}^{\mu_{2N-1}\mu_{2N}}+\frac
{1}{\ell_{\text{eff}(N)}^{2}}\delta_{\nu_{2N-1}\nu_{2N}}^{\mu_{2N-1}\mu_{2N}%
}\right)  =0,
\end{equation}
where the $\ell_{\text{eff}(i)}^{2}$ are the effective AdS radii corresponding to the vacua of the theory. If the
$i-$th vacuum is $\left(  k_{i}-1\right)  -$ degenerate, the corresponding
factor will appear $k_{i}$ terms in the factorization such that $k_{i}$ is
its algebraic multiplicity, and the sum of all algebraic multiplicities is
equal to $N$.

\subsection{EGB case}

In the particular case of EGB theory,  $\alpha_{0}$ and
$\alpha_{1}$ are given  in eq.(\ref{Einstein-AdS}), and of the higher
curvature couplings, only $\alpha_{2}\neq0$. Thus, we have%
\begin{align}
\Delta\left(  \lambda\right)   &  =%
{\displaystyle\sum\limits_{p=0}^{N}}
\frac{\alpha_{p}\left(  d-2\right)  !}{\left(  d-2p\right)  !}\left(
-1\right)  ^{p+1}\lambda^{p}\nonumber\\
&  =-\frac{1}{\ell^{2}}+\lambda-\left(  d-2\right)  \left(  d-3\right)
\alpha_{2}\lambda^{2}=0,
\end{align}
yielding
\begin{equation}
\lambda=\frac{-1\pm\sqrt{1-\frac{4\left(  d-2\right)  \left(  d-3\right)
\alpha_{2}}{\ell^{2}}}}{-2\left(  d-2\right)  \left(  d-3\right)  \alpha_{2}%
}=\frac{1\mp\sqrt{1-\frac{4\left(  d-2\right)  \left(  d-3\right)  \alpha_{2}%
}{\ell^{2}}}}{2\left(  d-2\right)  \left(  d-3\right)  \alpha_{2}},
\end{equation}
from which we have  
\begin{equation}
\ell_{\text{eff}\left(  \mp\right)  }^{2}=\frac{2\left(  d-2\right)  \left(
d-3\right)  \alpha_{2}}{1\mp\sqrt{1-\frac{4\left(  d-2\right)  \left(
d-3\right)  \alpha_{2}}{\ell^{2}}}}.
\end{equation}
Thus, the EOM factorizes as
\begin{equation}
E_{\nu}^{\mu}=\frac{\alpha_{2}}{2^{3}}\delta_{\nu\mu_{1}\mu_{2}\mu_{3}\mu_{4}%
}^{\mu\nu_{1}\nu_{2}\nu_{3}\nu_{4}}\left(  R_{\phantom{\mu_{1}\mu_{2}}\nu_{1}\nu_{2}}^{\mu_{1}\mu_{2}}
 +\frac{1}{\ell_{\text{eff}(-)}^{2}}\delta_{\nu_{1}\nu_{2}}^{\mu_{1}\mu_{1}%
}\right)  \left( R_{\phantom{\mu_{3}\mu_{4}}\nu_{3}\nu_{4}}^{\mu_{3}\mu_{4}} +\frac{1}{\ell
_{\text{eff}(+)}^{2}}\delta_{\nu_{3}\nu_{4}}^{\mu_{3}\mu_{4}}\right) \,.
\end{equation}
The degenerate case corresponds to%
\begin{equation}
\alpha_{2}=\frac{\ell^{2}}{4\left(  d-2\right)  \left(  d-3\right)  } \,,
\qquad  \ell_{\text{eff}\left(  \mp\right)  }^{2}=\frac{\ell^{2}}{2} \Rightarrow \Delta
^{\left(  1\right)  }=0,
\end{equation}
and so
\begin{equation}
E_{\nu}^{\mu}=\frac{\alpha_{2}}{2^{3}}\delta_{\nu\mu_{1}\mu_{2}\mu_{3}\mu_{4}%
}^{\mu\nu_{1}\nu_{2}\nu_{3}\nu_{4}}\left(  R_{\phantom{\mu_{1}\mu_{2}}\nu_{1}\nu_{2}}^{\mu_{1}\mu_{2}}
 +\frac{2}{\ell^{2}}\delta_{\nu_{1}\nu_{2}}^{\mu_{1}\mu_{1}%
}\right)  \left( R_{\phantom{\mu_{3}\mu_{4}}\nu_{3}\nu_{4}}^{\mu_{3}\mu_{4}} +\frac{2}{\ell^{2}}\delta_{\nu_{3}\nu_{4}}^{\mu_{3}\mu_{4}}\right)
 =0 \,.
\end{equation}

\section{$P(\mathcal{F})$ and degeneracy conditions}\label{AppendixC} 

We now study how the degeneracy conditions  introduced in eq.(\ref{k-thDeg}) appear in the coefficients of $P(  \mathcal{F}) $.
Recalling \eqref{KountertermCoupling} and \eqref{cd}, 
from eq.(\ref{P(F)BulkFixedJ}), the $\mathcal{F}^{2}$ coefficient is
given by%
\begin{align}
c_{2}  &  =%
{\displaystyle\sum\limits_{p=2}^{\frac{d+1}{2}}}
\frac{\left(  -1\right)  ^{p-2}p!\left(  d-3\right)  !\alpha_{p}}%
{2^{2}2!\left(  p-2\right)  !\left(  d+1-2p\right)  !\ell_{\text{eff}%
}^{2\left(  p-2\right)  }}\nonumber\\
&  =\left(
{\displaystyle\sum\limits_{p=2}^{\frac{d-1}{2}}}
\frac{\left(  -1\right)  ^{p-2}p!\left(  d-3\right)  !\alpha_{p}}%
{2^{2}2!\left(  p-2\right)  !\left(  d+1-2p\right)  !\ell_{\text{eff}%
}^{2\left(  p-2\right)  }} \right. \nonumber\\
& \qquad\qquad + \left.  \frac{\left(  -1\right)  ^{\left(  \frac{d+1}{2}\right)  -2}\left(
\frac{d+1}{2}\right)  !\left(  d-3\right)  !}{2^{2}2!\left(  \left(
\frac{d+1}{2}\right)  -2\right)  !\ell_{\text{eff}}^{2\left(  \left(
\frac{d+1}{2}\right)  -2\right)  }}\left(  \frac{2}{d+1}%
{\displaystyle\sum\limits_{p=1}^{\frac{d-1}{2}}}
\frac{p\alpha_{p}\left(  -1\right)  ^{\frac{d+3}{2}-p}}{\left(  d+1-2p\right)
!}\left(  \ell_{\text{eff}}\right)  ^{\left(  d+1-2p\right)  }\right)  \right)
\nonumber\\
&  =%
{\displaystyle\sum\limits_{p=2}^{\frac{d-1}{2}}}
\left(  \frac{\left(  -1\right)  ^{p}\left(  d-3\right)  !p\alpha_{p}}%
{2^{2}2!\left(  d+1-2p\right)  !\ell_{\text{eff}}^{2\left(  p-2\right)  }%
}\left(  \left(  p-1\right)  -\left(  \frac{d-1}{2}\right)  \right)  \right)
+\frac{\alpha_{1}}{2^{4}\left(  d-2\right)  }\left(  \ell_{\text{eff}}\right)
^{2}\nonumber\\
&  =%
{\displaystyle\sum\limits_{p=1}^{\frac{d-1}{2}}}
\left(  \frac{\left(  -1\right)  ^{p+1}\left(  d-3\right)  !p\alpha_{p}}%
{2^{4}\left(  d-2p\right)  !\ell_{\text{eff}}^{2\left(  p-2\right)  }}\right)
\nonumber\\
&  =\frac{\ell_{\text{eff}}^{2}}{2^{4}\left(  d-2\right)  }\Delta^{\left(
1\right)  }\left(  \ell_{\text{eff}}^{-2}\right)  . \label{c_2}%
\end{align}
Therefore, $\mathcal{F}^{2}$ term is proportional to the first degeneracy
condition.  Consequently --for degenerate points-- the Noether
prepotential is not linear but of higher order in $\mathcal{F}$, such that it is of
normalizable order (in agreement with the results of ref.\cite{Arenas-Henriquez:2019rph}.

Now, we consider the coefficient of the $\mathcal{F}^{3}$ term, which is
given by%
\begin{align}
c_{3}  &  =\sum\limits_{p=3}^{\frac{d+1}{2}}\frac{\left(  -1\right)
^{p-3}\left(  d-5\right)  !p!a_{p}}{2^{3}3!\left(  p-3\right)  !\left(
d+1-2p\right)  !\ell_{\text{eff}}^{2(p-3)}}\nonumber\\
&  =\left(  \sum\limits_{p=3}^{\frac{d-1}{2}}\frac{\left(  -1\right)
^{p-3}\left(  d-5\right)  !p!a_{p}}{2^{3}3!\left(  p-3\right)  !\left(
d+1-2p\right)  !\ell_{\text{eff}}^{2(p-3)}} \right. \nonumber\\
& \qquad\qquad  + \left.  \frac{\left(  -1\right)  ^{\frac{d-5}{2}}\left(  d-5\right)
!\binom{d+1}{2}!}{2^{3}3!\binom{d-5}{2}!\ell_{\text{eff}}^{d-5}}\frac{2}%
{d+1}\sum_{p=1}^{\frac{d-1}{2}}\frac{pa_{p}\left(  -1\right)  ^{\frac{d+3}%
{2}-p}}{\left(  d+1-2p\right)  !}\ell_{\text{eff}}^{d+1-2p}\right) \nonumber\\
&  =\left(  \sum\limits_{p=3}^{\frac{d-1}{2}}\frac{\left(  -1\right)
^{p}\left(  d-5\right)  !p!a_{p}}{2^{3}3!\left(  p-3\right)  !\left(
d+1-2p\right)  !\ell_{\text{eff}}^{2(p-3)}}\frac{\left(  d+1-2p\right)
\left(  d+2p-5\right)  }{4(p-1)\left(  p-2\right)  } \right. \nonumber\\
&\qquad\qquad +  \left.  \frac{\left(  d-5\right)  !\left(  d-1\right)  \left(  d-3\right)
}{2^{3}3!4\ell_{\text{eff}}^{d-5}}\left(  -\frac{a_{1}}{\left(  d-1\right)
!}\ell_{\text{eff}}^{d-1}+\frac{2a_{2}}{\left(  d-3\right)  !}\ell
_{\text{eff}}^{d-3}\right)  \right) \nonumber\\
&  =\sum\limits_{p=1}^{\frac{d-1}{2}}\frac{\left(  -1\right)  ^{p}\left(
d-5\right)  !p!\left(  d-3\right)  a_{p}}{2^{5}3!\left(  p-1\right)  !\left(
d-2p\right)  !\ell_{\text{eff}}^{2(p-2)}}\ell_{\text{eff}}^{2}+\sum
\limits_{p=2}^{\frac{d-1}{2}}\frac{\left(  -1\right)  ^{p}\left(  d-5\right)
!p!a_{p}}{2^{4}3!\left(  p-2\right)  !\left(  d-2p\right)  !\ell_{\text{eff}%
}^{2(p-2)}}\ell_{\text{eff}}^{2}\nonumber\\
&  =-\frac{\ell_{\text{eff}}^{4}}{2^{5}3!\left(  d-2\right)  \left(  d-4\right)
}\left(  \Delta^{\left(  1\right)
}+\frac{4}{\left(  d-3\right)\ell^{2}_{\text{eff}}}\Delta^{\left(  2\right)  }\right)  ,
\label{c_3}%
\end{align}
which we see is proportional to a linear combination of the first and second
degeneracy conditions.
Then, we consider the coefficient of the $\mathcal{F}^{4}$ term, which is given by
\begin{align}
c_{4}&=\sum\limits_{p=4}^{\frac{d+1}{2}}\frac{\left(  -1\right)  ^{p-4}%
p!\left(  d-7\right)  !\alpha_{p}}{2^{4}4!\left(  p-4\right)  !\left(
d+1-2p\right)  !\ell_{\text{eff}}^{2(p-4)}}\nonumber\\
&=\sum\limits_{p=4}^{\frac{d-1}{2}}\frac{\left(  -1\right)  ^{p-4}p!\left(
d-7\right)  !\alpha_{p}}{2^{4}4!\left(  p-4\right)  !\left(  d+1-2p\right)
!\ell_{\text{eff}}^{2(p-4)}}+\frac{(-1)^{\frac{d+1}{2}-4}\binom{d+1}{2}!\left(  d-7\right)  !}%
{2^{4}4!\left(  \frac{d+1}{2}-4\right)  !\ell_{\text{eff}}^{d-7}}\frac{2}%
{d+1}\sum\limits_{p=1}^{\frac{d-1}{2}}\frac{\left(  -1\right)  ^{\frac{d+3}%
{2}-p}p\alpha_{p}}{\left(  d+1-2p\right)  !}\ell_{\text{eff}}^{d+1-2p}%
\nonumber\\
&=\sum\limits_{p=4}^{\frac{d-1}{2}}\frac{\left(  -1\right)  ^{p-4}p!\left(
d-7\right)  !\alpha_{p}}{2^{4}4!\left(  p-4\right)  !\left(  d+1-2p\right)
!\ell_{\text{eff}}^{2(p-4)}}\left[  1-\frac{\left(  d-1\right)  \left(
d-3\right)  \left(  d-5\right)  }{2^{3}\left(  p-1\right)  \left(  p-2\right)
\left(  p-3\right)  }\right]  \nonumber\\
&-\frac{\left(  d-1\right)  \left(  d-3\right)  \left(  d-5\right)  }{2^{3}%
}\sum\limits_{p=1}^{3}\frac{\left(  -1\right)  ^{p-4}p!\left(  d-7\right)
!\alpha_{p}}{2^{4}4!\left(  p-1\right)  !\left(  d+1-2p\right)  !\ell
_{\text{eff}}^{2(p-4)}}.
\end{align}
The square bracket expression in the last equality can be written as
\begin{equation}
1-\frac{\left(  d-1\right)  \left(  d-3\right)  \left(  d-5\right)  }%
{2^{3}\left(  p-1\right)  \left(  p-2\right)  \left(  p-3\right)  }%
=-\frac{\left(  d-2p+1\right)  \left(  d^{2}-10d+33+4p^{2}+2dp-22p\right)
}{2^{3}\left(  p-1\right)  \left(  p-2\right)  \left(  p-3\right)  } \,,
\end{equation}
which upon substituted into the previous expression, one gets
\begin{align}
c_{4}&=
\frac{1}{2^{7}4!(d-2)\left(  d-4\right)  \left(  d-6\right)  }\bigg[\sum
\limits_{p=4}^{\frac{d-1}{2}}\frac{\left(  -1\right)  ^{p+1}p!\left(
d-2\right)  !\alpha_{p}}{\left(  p-4\right)  !\left(  d-2p\right)
!\ell_{\text{eff}}^{2(p-4)}}\frac{d^{2}-10d+33+4p^{2}+2dp-22p}{\left(
p-1\right)  \left(  p-2\right)  \left(  p-3\right)  \left(  d-3\right)
\left(  d-5\right)  }\nonumber\\
&+\sum\limits_{p=1}^{3}\frac{\left(  -1\right)  ^{p+1}p!\left(  d-2\right)
!\alpha_{p}}{\left(  p-1\right)  !\left(  d-2p\right)  !\ell_{\text{eff}%
}^{2(p-4)}}\frac{d-1}{d+1-2p}\bigg]\nonumber\\
&=\frac{1}{2^{7}4!(d-2)\left(  d-4\right)  \left(  d-6\right)  }\bigg[\ell
_{\text{eff}}^{6}\sum\limits_{p=1}^{\frac{d-1}{2}}\frac{\left(  -1\right)
^{p+1}p!\left(  d-2\right)  !\alpha_{p}}{\left(  p-1\right)  !\left(
d-2p\right)  !\ell_{\text{eff}}^{2(p-1)}}+\frac{4\ell_{\text{eff}}^{4}}%
{d-3}\sum\limits_{p=2}^{\frac{d-1}{2}}\frac{\left(  -1\right)  ^{p+1}p!\left(
d-2\right)  !\alpha_{p}}{2!\left(  p-2\right)  !\left(  d-2p\right)
!\ell_{\text{eff}}^{2(p-2)}}\nonumber\\
&+\frac{4!\ell_{\text{eff}}^{2}}{\left(  d-3\right)  \left(  d-5\right)  }%
\sum\limits_{p=3}^{\frac{d-1}{2}}\frac{\left(  -1\right)  ^{p+1}p!\left(
d-2\right)  !\alpha_{p}}{3!\left(  p-3\right)  !\left(  d-2p\right)
!\ell_{\text{eff}}^{2(p-3)}}\bigg]\nonumber\\
&=\frac{\ell_{\text{eff}}^{6}}{2^{7}4!(d-2)\left(  d-4\right)  \left(
d-6\right)  }\left[  \Delta^{\left(  1\right)  }+\frac{4}{\left(  d-3\right)
\ell_{\text{eff}}^{2}}\Delta^{\left(  2\right)  }+\frac{24}{\left(
d-3\right)  \left(  d-5\right)  \ell_{\text{eff}}^{4}}\Delta^{\left(
3\right)  }\right]  .
\end{align}

We infer from the previous coefficients that the generic term is given by
\begin{align}
c_{i\geq3}  & =\frac{\left(  -1\right)  ^{i}\ell_{\text{eff}}^{2i-2}}%
{2^{2i-1}i!%
{\displaystyle\prod\limits_{j=1}^{i-1}}
\left(  d-2j\right)  }\left(  \Delta^{\left(  1\right)  }+%
{\displaystyle\sum\limits_{l=2}^{i-1}}
\frac{2^{l-1} l!}{\left(
{\displaystyle\prod\limits_{m=1}^{l-1}}
\left(  d-1-2m\right)  \right)\ell_{\text{eff}}^{2l-2}  }\Delta^{\left(  l\right)  }\right)  
\label{gencoeffci}
\end{align}
which explicitly reproduces the coefficients up to and including   $c_{4}$.  Using Mathematica, for arbitrary $c_{i}$
we have verified \eqref{gencoeffci}   for the particular Lovelock Unique Vacuum theories considered in ref.\cite{Kastor:2006vw}. 
Finally,  since
\begin{equation}
c_{i}=%
{\displaystyle\sum\limits_{l=1}^{i-1}}
p_{\left(  i,l\right)  }\Delta^{\left(  l\right)  },
\end{equation}
we have%
\begin{align}
p_{\left(  2,1\right)  }  & =\frac{\ell_{\text{eff}}^{2}}{2^{3}2!\left(
d-2\right)  } \qquad 
p_{\left(  i\geq3,l=1\right)  }   =\frac{\left(  -1\right)  ^{i}%
\ell_{\text{eff}}^{2i-2}}{2^{2i-1}i!%
{\displaystyle\prod\limits_{j=1}^{i-1}}
\left(  d-2j\right)  } \nonumber\\
p_{\left(  i\geq3,l\geq2\right)  }  & =\frac{\left(  -1\right)  ^{i}%
\ell_{\text{eff}}^{2\left(  i-l\right)  }2^{l-2i}l!}{i!\left(
{\displaystyle\prod\limits_{j=1}^{i-1}}
\left(  d-2j\right)  \right)  \left(
{\displaystyle\prod\limits_{m=1}^{l-1}}
\left(  d-1-2m\right)  \right)  }.%
\label{P_ij}
\end{align}

\section{Noether prepotential from $P(
\mathcal{F} )  $}\label{AppendixD}

The Noether prepotential can be obtained by computing the on-shell variation
of the bulk gravity action.  This
prepotential is important for computing the asymptotic charges of the theory
\cite{Kofinas:2007ns,Arenas-Henriquez:2019rph},
as it is a covariant charge density that has to be integrated over co-dimension 2
surfaces at infinity.

Starting from the bulk $P\left(  \mathcal{F}\right)  $ defined in
eq.(\ref{P(F)BulkFixedJ}) 
\begin{equation}
P_{\left(  d+1\right)  ,\left\{  \alpha_{p}\right\}  }\left(  \mathcal{F}%
\right)  =%
{\displaystyle\sum\limits_{j=2}^{\frac{d+1}{2}}}
c_{j}\,\delta_{\mu_{1}\cdots\mu_{2j}}^{\nu_{1}\cdots\nu_{2j}}\mathcal{F}%
_{\nu_{1}\nu_{2}}^{\mu_{1}\mu_{2}}\cdots\mathcal{F}_{\nu_{2j-1}\nu_{2j}}%
^{\mu_{2j-1}\mu_{2j}} \,,
\end{equation}
where   $c_{j}$ is given in eq.(\ref{Cj}), we have
\begin{equation}
I_{L}=\frac{1}{16\pi G}%
{\displaystyle\int\limits_{M}}
d^{d+1}x\sqrt{-\mathcal{G}}\,P_{\left(  d+1\right)  ,\left\{  \alpha
_{p}\right\}  }\left(  \mathcal{F}\right)  +\tau_{d}\chi\left(  M\right)  ,
\end{equation}
for the renormalized Lovelock-AdS action (for odd $d$). Its on-shell variation is
\begin{equation}
\delta I_{L}=\frac{1}{16\pi G}%
{\displaystyle\int\limits_{\partial M}}
d^{d}x\sqrt{-h}%
{\displaystyle\sum\limits_{j=2}^{\frac{d+1}{2}}}
\left(  -2jc_{j}\right)  \delta_{i_{1}\cdots i_{2j-1}}^{k_{1}\cdots k_{2j-1}%
}\mathcal{F}_{k_{1}k_{2}}^{i_{1}i_{2}}\mathcal{\cdots F}_{k_{2j-3}k_{2j-2}%
}^{i_{2j-3}i_{2j-2}}\left(  \left(  h^{-1}\delta h\right)  _{l}^{i_{2j-1}%
}K_{k_{2j-1}}^{l}+2\delta K_{k_{2j-1}}^{i_{2j-1}}\right)  ,
\label{LovelockOnshellVariation}%
\end{equation}
neglecting boundary terms at the AdS boundary. Therefore, in the case of
non-degenerate theories,  assuming  asymptotic conformal flatness \cite{Anastasiou:2019ldc} and
considering that $\mathcal{F}$ has the fall-off of the normalizable mode, we
have that at the normalizable order
\begin{align}
\delta I_{L}  &  =\frac{1}{16\pi G}%
{\displaystyle\int\limits_{\partial M}}
d^{d}x\sqrt{-h}\left( -4c_{2}\right)  \delta
_{j_{1}j_{2}j_{3}}^{i_{1}i_{2}i_{3}}\mathcal{F}_{i_{1}i_{2}}^{j_{1}j_{2}%
}\left(  h^{-1}\delta h\right)^{j_{3}}_{l} K_{i_{3}}^{l}\,,
\nonumber\\
&  =\frac{1}{16\pi G}%
{\displaystyle\int\limits_{\partial M}}
d^{d}x\sqrt{-h}\left(  \frac{-16c_{2}}{\ell_{\text{eff}}}\right)  \mathcal{E}
_{j}^{i}\left(  h^{-1}\delta h\right)  _{i}^{j},
\end{align}
where we have taken just the leading order in the FG expansion of the extrinsic curvature $K_{j_{3}}^{l} =\delta_{j_{3}}^{l}/\ell_{\text{eff}}$, the electric part of
the Weyl tensor $\mathcal{E}_{j}^{i}=\left(  -W_{jl}^{il}\right)  $ and the fact that $W\simeq\mathcal{F}$ up to the normalizable order.

Finally, noting from eq.(\ref{c_2}) that%
\begin{equation}
c_{2}=%
{\displaystyle\sum\limits_{p=2}^{\frac{d+1}{2}}}
\frac{\left(  -1\right)  ^{p}p!\left(  d-3\right)  !\alpha_{p}}{8\left(
p-2\right)  !\left(  d+1-2p\right)  !\ell_{\text{eff}}^{2\left(  p-2\right)
}}=\frac{\ell_{\text{eff}}^{2}}{2^{4}\left(  d-2\right)  }\Delta^{\left(
1\right)  }\left(  \ell_{\text{eff}}^{-2}\right)  ,
\end{equation}
with $\Delta^{\left(  1\right)  }$ defined as in eq.(\ref{k-thDeg}), we have
\begin{equation}
\delta I_{L}=-\frac{\ell_{\text{eff}}}{16\pi G\left(
d-2\right)  }\Delta^{\left(  1\right)  }\left(  \ell_{\text{eff}}^{-2}\right)
%
{\displaystyle\int\limits_{\partial M}}
d^{d}x\sqrt{-h}\,\mathcal{E}_{j}^{i}\left(  h^{-1}\delta h\right)  _{i}^{j} \,,
\end{equation}
from which we can directly read the Noether prepotential $\tau_{j}^{i}$ 
\begin{equation}
\delta I_{L}=%
{\displaystyle\int\limits_{\partial M}}
d^{d}x\sqrt{-h}\left(  \frac{1}{2}\tau_{j}^{i}\right)  \left(  h^{-1}\delta
h\right)  _{i}^{j},
\end{equation}
such that%
\begin{equation}
\tau_{i}^{j}=-\frac{\ell_{\text{eff}}}{8\pi G\left(  d-2\right)  }%
\Delta^{\left(  1\right)  }\left(  \ell_{\text{eff}}^{-2}\right)  
\mathcal{E}_{j}^{i} \,,
\end{equation}
in agreement with ref.\cite{Arenas-Henriquez:2019rph}. As $\Delta^{\left(  1\right)  }\left(
\ell_{\text{eff}}^{-2}\right)  =1$ for Einstein-AdS, this reproduces the known
result for the Noether prepotential in this case.

\subsection{Degenerate case}

In the first-degenerate case, we have that $\Delta^{\left(  1\right)  }\left(
\ell_{\text{eff}}^{-2}\right)  =0$, implying $c_{2}=0$, and $\Delta^{\left(  k>1\right)  }\left(
\ell_{\text{eff}}^{-2}\right)  \neq0$. From
eq.(\ref{c_3})  
\begin{align}
c_{3}  &  =-\frac{\ell_{\text{eff}}^{2}}{\left(  d-2\right)  \left(
d-4\right)  2^{4}3}\left(  \frac{1}{\left(  d-3\right)  }\Delta^{\left(
2\right)  }+\frac{\ell_{\text{eff}}^{2}}{4}\Delta^{\left(  1\right)  }\right)
\nonumber\\
&  =-\frac{\ell_{\text{eff}}^{2}}{\left(  d-2\right)  \left(  d-3\right)
\left(  d-4\right)  2^{4}3}\Delta^{\left(  2\right)  }\left(  \ell
_{\text{eff}}^{-2}\right) \,,
\end{align}
and so the eq.\eqref{LovelockOnshellVariation} can be written to the lowest order as
\begin{equation}
\delta I_{L}=\left(  \frac{6\ell_{\text{eff}}\Delta^{\left(
2\right)  }\left(  \ell_{\text{eff}}^{-2}\right)  }{16\pi G\left(  d-2\right)
\left(  d-3\right)  \left(  d-4\right)  2^{4}3}\right)
{\displaystyle\int\limits_{\partial M}}
d^{d}x\sqrt{-h}\delta_{j_{1}\cdots j_{5}}^{i_{1}\cdots i_{5}}\mathcal{F}%
_{i_{1}i_{2}}^{j_{1}j_{2}}\mathcal{F}_{i_{3}i_{4}}^{j_{3}j_{4}}\left(
h^{-1}\delta h\right)  _{i_{5}}^{j_{5}} \,.
\end{equation}
Here, it is clear that the Noether prepotential is proportional to an
antisymmetric contraction along boundary indices of $\mathcal{F}^{2}$. This is
in agreement with ref.\cite{Arenas-Henriquez:2019rph}, where the asymptotic fall-off of said
contraction was shown to be precisely the normalizable mode, making the Noether
prepotential finite.

In the $k-$degenerate case, we have $\Delta^{\left(  q\leq k\right)  }=0$,
$\Delta^{\left(  q>k\right)  }\neq0$, and because $c_{q}=%
{\displaystyle\sum\limits_{i=1}^{q-1}}
p_{(q,i)}\Delta^{\left(  i\right)  }$ for some $p_{(q,i)}$ coefficients (as
shown in eq.\eqref{P_ij}), we have that $c_{q\leq k+1}=0$ and
$c_{k+2}=p_{\left(  k+2,k+1\right)  }\Delta^{\left(  k+1\right)  }$. Thus, to
lowest order we get
\begin{align}
\delta I_{L} &  =\left(  \frac{-2\left(  k+2\right)  c_{k+2}%
}{16\pi G\ell_{\text{eff}}}\right)
{\displaystyle\int\limits_{\partial M}}
d^{d}x\sqrt{-h}\delta_{j_{1}\cdots j_{2k+3}}^{i_{1}\cdots i_{2k+3}}%
\mathcal{F}_{i_{1}i_{2}}^{j_{1}j_{2}}\cdots\mathcal{F}_{i_{2k+1}i_{2k+2}%
}^{j_{2k+1}j_{2k+2}}\left(  h^{-1}\delta h\right)  _{i_{2k+3}}^{j_{2k+3}%
},\nonumber\\
&  =\left(  \frac{-2\left(  k+2\right)  p_{\left(  k+2,k+1\right)  }%
\Delta^{\left(  k+1\right)  }\left(  \ell_{\text{eff}}^{-2}\right)  }{16\pi
G\ell_{\text{eff}}}\right)
{\displaystyle\int\limits_{\partial M}}
d^{d}x\sqrt{-h}\delta_{j_{1}\cdots j_{2k+3}}^{i_{1}\cdots i_{2k+3}}%
\mathcal{F}_{i_{1}i_{2}}^{j_{1}j_{2}}\cdots\mathcal{F}_{i_{2k+1}i_{2k+2}%
}^{j_{2k+1}j_{2k+2}}\left(  h^{-1}\delta h\right)  _{i_{2k+3}}^{j_{2k+3}} \,.
\label{deltaIonshell}
\end{align}
Note that the Noether prepotential is proportional to an
antisymmetric contraction along boundary indices of $\mathcal{F}^{k+1}$, in
agreement with
ref.\cite{Arenas-Henriquez:2019rph}. There it is shown that the eq.\eqref{deltaIonshell} falls-off asymptotically as the normalizable mode, thus rendering it finite. Direct comparison with ref. \cite{Arenas-Henriquez:2019rph} fixes the $p_{\left(k+2,k+1\right)  }$ coefficient to be that in eq.\eqref{DegP}, such that for a
$k$-degenerate theory, the lowest order coefficient in the bulk $P\left(
\mathcal{F}\right)  $ is given by%
\begin{equation}
c_{k+2}=p_{\left(  k+2,k+1\right)  }\Delta^{\left(  k+1\right)  }\left(
\ell_{\text{eff}}^{-2}\right)  .\label{DegCoef}%
\end{equation}

\section{ JM extremal surface for ball-shaped
entangling region}\label{AppendixE}

We proceed to verify that the spherical hemisphere is indeed the extremal
surface of the JM functional for the case of ball-shaped entangling regions in
pure AdS (dual to the ground state of a CFT in Minkowski spacetime). In order
to see this, we consider that the bulk metric of the Poincar\'{e} patch of
pure AdS (for a particular vacuum characterized by $\ell_{\text{eff}}^{2}$),
is written in the FG gauge as%
\begin{align}
ds_{G}^{2}  &  =G_{\mu\nu}dx^{\mu}dx^{\nu}=\frac{\ell_{\text{eff}}^{2}%
d\rho^{2}}{4\rho^{2}}+\frac{-dt^{2}+dr^{2}+r^{2}d\Omega_{d-2}^{2}}{ \rho^2
},\nonumber\\
d\Omega_{d-2}^{2}  &  =d\theta_{1}^{2}+\sin^{2}\theta_{1}d\theta_{1}%
^{2}+\cdots+\sin^{2}\theta_{1}\cdots\sin^{2}\theta_{d-3}^{2}d\theta_{d-2}^{2}.
\end{align}
Consider 
the spherical hemisphere  
\begin{equation}
\Sigma:\left\{  t=\text{const. };\text{ }r=\sqrt{R^{2}-\ell_{\text{eff}}%
^{2}\rho}\right\}  , \label{AnsatzSigma}%
\end{equation}
with $\left\{ y^{\alpha
}\right\}  =\left\{  \rho,\theta_{1},\ldots,\theta_{d-2}\right\}  $ being the worldvolume coordinates of $\Sigma$. The
intrinsic metric 
\begin{equation}
\gamma_{\alpha\beta}=\frac{\partial x^{\mu}}{\partial y^{\alpha}}%
\frac{\partial x^{\nu}}{\partial y^{\beta}}G_{\mu\nu},
\end{equation}
on $\Sigma$ is then
\begin{equation}
ds_{\gamma}^{2}=\gamma_{\alpha\beta}dy^{\alpha}dy^{\beta}=\frac{\ell
_{\text{eff}}^{2}}{4\rho^{2}}\left(  1+\frac{\ell_{\text{eff}}^{2}\rho}%
{R^{2}-\ell_{\text{eff}}^{2}\rho}\right)  d\rho^{2}+\frac{\left(  R^{2}%
-\ell_{\text{eff}}^{2}\rho\right)  }{\rho}d\Omega_{d-2}^{2} \,.
\label{IntrinsicMetricHemisphere}%
\end{equation}
In order to check that the ansatz (\ref{AnsatzSigma}) for the surface $\Sigma$  indeed minimizes the JM functional, we evaluate the corresponding equations of motion \eqref{JMEOM}.  Since  $\Sigma$ is a constant curvature surface, we have
\begin{equation}
\widehat{\mathcal{R}}_{\phantom{\alpha_{1}\alpha_{2}}\beta_{1}\beta_{2}}^{\alpha_{1}\alpha_{2}} [\gamma]=-\frac{1}%
{\ell_{\text{eff}}^{2}}\delta_{\beta_{1}\beta_{2}}^{\alpha_{1}\alpha_{2}},
\end{equation}
for its Riemann curvature tensor. From eq.\eqref{JMEOM} we have%
\begin{align}
E_{JM}  &  =%
{\displaystyle\sum\limits_{p=0}^{\left\lfloor \frac{d-2}{2}\right\rfloor }}
\frac{\left(  -1\right)  ^{p}\alpha_{\left(  p+1\right)  }\left(  p+1\right)
}{2^{p}\ell_{\text{eff}}^{2p}}^{p}\delta_{\beta_{1}\cdots\beta_{2p+1}}%
^{\gamma_{1}\cdots\gamma_{2p+1}}\mathcal{K}_{\gamma_{1}}^{\beta_{1}}%
\delta_{\gamma_{2}\gamma_{3}}^{\beta_{2}\beta_{3}}\cdots\delta_{\gamma
_{2p}\gamma_{2p+1}}^{\beta_{2p}\beta_{2p+1}}\nonumber\\
&  =%
{\displaystyle\sum\limits_{p=0}^{\left\lfloor \frac{d-2}{2}\right\rfloor }}
\frac{\left(  -1\right)  ^{p}\alpha_{\left(  p+1\right)  }\left(  p+1\right)
}{\ell_{\text{eff}}^{2p}}^{p}\delta_{\beta_{1}\cdots\beta_{2p+1}}^{\gamma
_{1}\cdots\gamma_{2p+1}}\delta_{\gamma_{1}}^{\beta_{1}}\cdots\delta
_{\gamma_{2p}}^{\beta_{2p}}\mathcal{K}_{\gamma_{2p+1}}^{\beta_{2p+1}%
}\nonumber\\
&  =\left(
{\displaystyle\sum\limits_{p=0}^{\left\lfloor \frac{d-2}{2}\right\rfloor }}
\frac{\left(  -1\right)  ^{p}\left(  d-2\right)  !\left(  p+1\right)
\alpha_{\left(  p+1\right)  }}{\left(  d-2-2p\right)  !\ell_{\text{eff}}^{2p}%
}\right)  tr\left[  \mathcal{K}\right]  ,
\end{align}
using eqs.\eqref{DissolveDeltaIdentity} and \eqref{ContractDeltaIdentity} from Appendix~\ref{AppendixA}.
Now, as shown in refs. \cite{Anastasiou:2020smm,Bhattacharyya:2014yga}, the spherical hemisphere trivially satisfies
$tr\left[  \mathcal{K}\right]  =0$ as it is a minimal surface in the
mathematical sense that it minimizes the area functional. In turn, this is equivalent to the statement that the RT surface is the correct one for the computation
of HEE in Einstein-AdS gravity). Thus, the spherical hemisphere also satisfies
$E_{JM}\left[  \Sigma\right]  =0$, and therefore it is the correct surface
that minimizes the JM functional for the case of a ball-shaped entangling
region in the CFT.

\section{Evaluation of the JM functional and co-dimension 2
Kounterterm}\label{AppendixF}

Here we evaluate the JM functional (\ref{I_JM_sum})  and the co-dimension 2 Kounterterm used in the
cancelation of divergences for the HEE. From  eq.\eqref{I_JM_decomp}, the JM functional can be written as the sum of three terms, such that
$$
4GS_{JM}\overset{\text{def.}}{=}I_{JM}=I_{JM}^{\left(  1\right)  }%
+I_{JM}^{\left(  2\right)  }+I_{JM}^{\left(  3\right)  } \,,
$$
that we now proceed to
evaluate. In order to simplify these terms,
we  employ the radial decomposition of the Riemann tensor of $\gamma
_{\alpha\beta}$ given in eq.\eqref{RiemannDecomp}, and the FG-like expansions \eqref{FG-likeMetric} and \eqref{FG-likeTensors}.

We start simplifying $I_{JM}^{\left(  1\right)  }$, where we have  
\begin{align}
I_{JM}^{\left(  1\right)  }  &  =%
{\displaystyle\int\limits_{\Sigma}}
d^{d-1}y\sqrt{\gamma}\sum\limits_{p=1}^{\left\lfloor \frac{d-2}{2}%
\right\rfloor }\frac{\left(  p+1\right)  p\alpha_{\left(  p+1\right)  }%
}{2^{p-2}}\delta_{b_{1}\ldots b_{2p-1}}^{a_{1}\ldots a_{2p-1}}\left(  \frac
{1}{N}\partial_{\rho}k_{a_{1}}^{b_{1}}-k_{c}^{b_{1}}k_{a_{1}}^{c}\right)
\nonumber\\
&\qquad \times  \left(  \mathcal{R}_{\phantom{b_{2}b_{3}}a_{2}a_{3}}^{b_{2}b_{3}}-2k_{a_{2}}^{b_{2}}k_{a_{3}%
}^{b_{3}}\right)  \cdots\left(  \mathcal{R}_{\phantom{b_{2p-2}b_{2p-1}}a_{2p-2}a_{2p-1}}^{b_{2p-2}b_{2p-1}}-2k_{a_{2p-2}}^{b_{2p-2}}k_{a_{2p-1}}^{b_{2p-1}}\right) \nonumber\\
&  =%
{\displaystyle\int\limits_{\Sigma}}
d^{d-1}y\sqrt{\gamma}\sum\limits_{p=1}^{\left\lfloor \frac{d-2}{2}%
\right\rfloor }\sum\limits_{m=0}^{p-1}\sum\limits_{s=0}^{2\left(
p-1-m\right)  }\frac{\left(  -1\right)  ^{p-m}\left(  p+1\right)  !\left(
2p-2m-2\right)  !\left(  d-2p+s\right)  !\alpha_{\left(  p+1\right)  }%
}{2^{m-1}m!s!\left(  p-1-m\right)  !\left(  2p-2m-2-s\right)  !\left(
d-1-2p\right)  !\ell_{\text{eff}}^{s+2}} \nonumber\\
&\qquad \times  \rho^{2p-m-s-2}\delta_{b_{1}\cdots b_{2p-s-2}}^{a_{1}\cdots a_{2p-s-2}%
}\mathcal{R}_{\phantom{b_{1}b_{2}}a_{1}a_{2}}^{b_{1}b_{2}}\left[  \sigma\right]  \cdots
\mathcal{R}_{\phantom{b_{2m-1}b_{2m}}a_{2m-1}a_{2m}}^{b_{2m-1}b_{2m}}\left[  \sigma\right]  \left(
k^{\left(  2\right)  }\right)  _{a_{2m+1}}^{b_{2m+1}}\cdots\left(  k^{\left(
2\right)  }\right)  _{a_{2p-s-2}}^{b_{2p-s-2}}+ \cdots \,,
\end{align}
where the ellipsis indicates higher order terms and $\mathcal{R}_{\phantom{a_{1}a_{2}}b_{1}b_{2}}%
^{a_{1}a_{2}}\left[  \sigma\right]  $ indicates the Riemann tensor of the
$\sigma_{ab}$ metric. Finally, we isolate the leading and next-to-leading
order divergences, to obtain%
\begin{align}
I_{JM}^{\left(  1\right)  }  &  =%
{\displaystyle\int\limits_{\Sigma}}
d^{d-1}y\sqrt{\gamma}\sum\limits_{p=1}^{\left\lfloor \frac{d-2}{2}%
\right\rfloor }\frac{\left(  -1\right)  ^{p}2\left(  p+1\right)
p\alpha_{\left(  p+1\right)  }}{\left(  d-1-2p\right)  !}\left(  \frac{\left(
d-2\right)  !}{\ell_{\text{eff}}^{2p}}+2\left(  p-1\right)  \rho\frac{\left(
d-3\right)  !}{\ell_{eff}^{2p-1}}tr\left[  k_{\left(  2\right)  }\right]
\right. \nonumber\\
& \qquad\qquad\phantom{ \frac{\left(
d-2\right)  !}{\ell_{\text{eff}}^{2p}}} \phantom{ \frac{\left(
d-2\right)  !}{\ell_{\text{eff}}^{2p}}} \phantom{ \frac{\left(
d-2\right)  !}{\ell_{\text{eff}}^{2p}}} \phantom{ \frac{\left(
d-2\right)  !}{\ell_{\text{eff}}^{2p}}}   - \left.  \left(  p-1\right)  \rho\frac{\left(  d-4\right)  !}{\ell
_{\text{eff}}^{2p-2}}\mathcal{R}\left[  \sigma\right]  \right)  + \cdots \,.
\label{F2}
\end{align}
\qquad
Considering the next term $I_{JM}^{\left(  2\right)  }$,  we can write 
\begin{gather}
I_{JM}^{\left(  2\right)  }=%
{\displaystyle\int\limits_{\Sigma}}
d^{d-1}y\sqrt{\gamma}\sum\limits_{p=2}^{\left\lfloor \frac{d-2}{2}%
\right\rfloor }\frac{\left(  p+1\right)  \left(  p-1\right)  p\alpha_{\left(
p+1\right)  }}{2^{p-2}}\delta_{b_{1}\ldots b_{2p-1}}^{a_{1}\ldots a_{2p-1}%
}\widehat{\mathcal{R}}_{\phantom{b_{2}b_{3}}a_{1}a_{2}}^{b_{1}\rho}\widehat{\mathcal{R}}%
_{\phantom{b_{2}b_{3}}a_{3}\rho}^{b_{2}b_{3}}\widehat{\mathcal{R}}_{\phantom{b_{2}b_{3}}a_{4}a_{5}}^{b_{4}b_{5}}%
\cdots\widehat{\mathcal{R}}_{\phantom{b_{2}b_{3}}a_{2p-2}a_{2p-1}}^{b_{2p-2}b_{2p-1}}\nonumber\\
=%
{\displaystyle\int\limits_{\Sigma}}
d^{d-1}y\sqrt{\gamma}\sum\limits_{p=2}^{\left\lfloor \frac{d-2}{2}%
\right\rfloor }\frac{\left(  p+1\right)  \left(  p-1\right)  p\alpha_{\left(
p+1\right)  }}{2^{p-4}}\delta_{b_{1}\ldots b_{2p-1}}^{a_{1}\ldots a_{2p-1}%
}\nabla_{a_{1}}k_{a_{2}}^{b_{1}}\nabla^{b_{2}}k_{a_{3}}^{b_{3}}\left(
\mathcal{R}_{\phantom{b_{2}b_{3}}a_{4}a_{5}}^{b_{4}b_{5}}-2k_{a_{4}}^{b_{4}}k_{a_{5}}^{b_{5}%
}\right)   \nonumber\\
\times \cdots\left(  \mathcal{R}_{a_{2p-2}a_{2p-1}}^{b_{2p-2}b_{2p-1}}-2k_{a_{2p-2}%
}^{b_{2p-2}}k_{a_{2p-1}}^{b_{2p-1}}\right)  \,.
\end{gather}

From the expansion of the extrinsic curvature $k_{ab}$ in eq.\eqref{FG-likeTensors}, the term $\nabla_{a_{1}}k_{a_{2}}^{b_{1}}$ is
$O\left(  \rho\right)  $, and  $\nabla^{b_{2}}k_{a_{3}}^{b_{3}}$ is $O\left(
\rho^{2}\right)  $.  Hence $\nabla_{a_{1}}k_{a_{2}}^{b_{1}}\nabla^{b_{2}%
}k_{a_{3}}^{b_{3}}$ is already $O\left(  \rho^{3}\right)  $, implying  $I_{JM}^{\left(  2\right)  }$ does not contribute up to the
next-to-leading divergence in $\rho$ and so can be neglected.

Finally, using eqs.(\ref{FG-likeMetric}-\ref{FG-likeTensors}) we can simplify $I_{JM}^{\left(  3\right)  }$ as
\begin{align}
I_{JM}^{\left(  3\right)  }  &  =%
{\displaystyle\int\limits_{\Sigma}}
d^{d-1}y\sqrt{\gamma}\sum\limits_{p=0}^{\left\lfloor \frac{d-2}{2}%
\right\rfloor }\frac{\left(  p+1\right)  \alpha_{\left(  p+1\right)  }}{2^{p}%
}\delta_{b_{1}\ldots b_{2p}}^{a_{1}\ldots a_{2p}}\left(  \mathcal{R}%
_{\phantom{b_{1}b_{2}}a_{1}a_{2}}^{b_{1}b_{2}}-2k_{a_{1}}^{b_{1}}k_{a_{2}}^{b_{2}}\right)
\cdots\left(  \mathcal{R}_{\phantom{b_{1}b_{2}}a_{2p-1}a_{2p}}^{b_{2p-1}b_{2p}}-2k_{a_{2p-1}%
}^{b_{2p-1}}k_{a_{2p}}^{b_{2p}}\right) \nonumber\\
&  =%
{\displaystyle\int\limits_{\Sigma}}
d^{d-1}y\sqrt{\gamma}\sum\limits_{m=0}^{\left\lfloor \frac{d-2}{2}%
\right\rfloor }\sum\limits_{p=m}^{\left\lfloor \frac{d-2}{2}\right\rfloor
}\sum\limits_{s=0}^{2\left(  p-m\right)  }\left(  -1\right)  ^{p-m}%
\frac{\left(  p+1\right)  \alpha_{\left(  p+1\right)  }p!\left(  2p-2m\right)
!\left(  d-2-2p+s\right)  !}{2^{m}\ell_{\text{eff}}^{s}m!\left(  p-m!\right)
s!\left(  2p-2m-s\right)  !\left(  d-2-2p\right)  !}\nonumber\\
&\qquad \times  \rho^{2p-m-s}\delta_{b_{1}\cdots b_{2p-1}}^{a_{1}\cdots a_{2p-s}%
}\mathcal{R}_{\phantom{b_{1}b_{2}}a_{1}a_{2}}^{b_{1}b_{2}}\left[  \sigma\right]  \cdots
\mathcal{R}_{\phantom{b_{1}b_{2}}a_{2m-1}a_{2m}}^{b_{2m-1}b_{2m}}\left[  \sigma\right]  \left(
k^{\left(  2\right)  }\right)  _{a_{2m+1}}^{b_{2m+1}}\cdots\left(  k^{\left(
2\right)  }\right)  _{a_{2p-s}}^{b_{2p-s}} + \cdots \,.
\end{align}
Isolating the leading and next-to-leading orders in $\rho$ yields
\begin{align}
I_{JM}^{\left(  3\right)  }  &  =%
{\displaystyle\int\limits_{\Sigma}}
d^{d-1}y\sqrt{\gamma}\sum\limits_{p=0}^{\left\lfloor \frac{d-2}{2}%
\right\rfloor }\left(  -1\right)  ^{p}\frac{\left(  p+1\right)  \alpha
_{\left(  p+1\right)  }}{\left(  d-2-2p\right)  !}\left(  \frac{\left(
d-2\right)  !}{\ell_{\text{eff}}^{2p}}-\rho p\left(  \frac{\left(  d-4\right)
!}{\ell_{\text{eff}}^{2p-2}}\mathcal{R}\left[  \sigma\right]   \right.
\right. \nonumber\\
&\qquad\qquad\qquad\qquad\qquad\qquad +  \left.  \left.  2\frac{\left(  d-3\right)  !}{\ell_{\text{eff}}^{2p}%
}\left(  tr\left[  \sigma^{\left(  2\right)  }\right]  +\frac{\ell
_{\text{eff}}^{2}\kappa_{a}^{\left(  i\right)  a}\kappa_{b}^{\left(  i\right)
b}}{2(d-2)}\right)  \right)  \right)  +  \cdots \,,
\label{F5}
\end{align}
while implementing eq.\eqref{F2} along with \eqref{F5} results in eq.\eqref{I_JM_sum} for the  JM functional $I_{JM}=I_{JM}^{\left(  1\right)  }+I_{JM}^{\left(  3\right)  }$  up to the next-to-leading divergent order.

Now, we focus on the analysis of  the Kounterterm $I_{KT}$. For odd $d$, we have%
\begin{align}
I_{KT}^{\operatorname*{odd}}  &  =4GS_{KT}^{\operatorname*{odd}}=\left\lfloor
\frac{d+1}{2}\right\rfloor c_{d}\int_{\partial\Sigma} d^{d-2}yB_{d-2}\nonumber\\
&  =\left(  \frac{d+1}{2}\right)  c_{d}\left(  -\left(  d-1\right)  \right)
{\displaystyle\int\limits_{\partial\Sigma}}
d^{d-2}y\sqrt{\tilde{\gamma}} {\displaystyle\int\limits_{0}^{1}}
dt\delta_{b_{1}\cdots b_{d-2}}^{a_{1}\cdots a_{d-2}}k_{a_{1}}^{b_{1}}\left(
\frac{1}{2}\mathcal{R}_{\phantom{b_{2}b_{3}}a_{2}a_{3}}^{b_{2}b_{3}}-t^{2}k_{a_{2}}^{b_{2}%
}k_{a_{3}}^{b_{3}}\right) \nonumber\\
& \qquad \qquad  \qquad  \qquad  \qquad  \qquad  \qquad  \qquad  \times
  \cdots\left(  \frac{1}{2}\mathcal{R}_{\phantom{b_{2}b_{3}}a_{d-3}%
a_{d-2}}^{b_{d-3}b_{d-2}}-t^{2}k_{a_{d-3}}^{b_{d-3}}k_{a_{d-2}}^{b_{d-2}%
}\right) 
\end{align}
where the parametric integral ($PI$) can be expanded as
\begin{align}
PI  &  =%
{\displaystyle\int\limits_{0}^{1}}
dt\delta_{b_{1}\cdots b_{d-2}}^{a_{1}\cdots a_{d-2}}k_{a_{1}}^{b_{1}}\left(
\frac{1}{2}\mathcal{R}_{\phantom{b_{2}b_{3}}a_{2}a_{3}}^{b_{2}b_{3}}-t^{2}k_{a_{2}}^{b_{2}%
}k_{a_{3}}^{b_{3}}\right)  \cdots\left(  \frac{1}{2}\mathcal{R}_{\phantom{b_{2}b_{3}}a_{d-3}%
a_{d-2}}^{b_{d-3}b_{d-2}}-t^{2}k_{a_{d-3}}^{b_{d-3}}k_{a_{d-2}}^{b_{d-2}%
}\right) \nonumber\\
&  =\sum\limits_{m=0}^{\frac{d-3}{2}}\frac{\left(  \frac{d-3}{2}\right)
!\left(  -1\right)  ^{\frac{d-3}{2}-m}}{m!\left(  \frac{d-3}{2}-m\right)
!2^{m}}\left(
{\displaystyle\int\limits_{0}^{1}}
dtt^{d-3-2m}\right)  \delta_{b_{1}\cdots b_{d-2}}^{a_{1}\cdots a_{d-2}%
}\mathcal{R}_{\phantom{b_{2}b_{3}}a_{1}a_{2}}^{b_{1}b_{2}}\cdots\mathcal{R}_{\phantom{b_{2}b_{3}}a_{2m-1}a_{2m}%
}^{b_{2m-1}b_{2m}}k_{a_{2m+1}}^{b_{2m+1}}\cdots k_{a_{d-2}}^{b_{d-2}%
}\nonumber\\
&  =\sum\limits_{m=0}^{\frac{d-3}{2}}\sum\limits_{s=0}^{d-2-2m}\frac{\left(
-1\right)  ^{\frac{d-3}{2}-m}\left(  \frac{d-3}{2}\right)  !\left(
d-3-2m\right)  !\epsilon^{d-2-m-s}}{m!\left(  \frac{d-3}{2}-m\right)  !\left(
d-2-2m-s\right)  !2^{m}\ell_{\text{eff}}^{s}}\delta_{b_{1}\cdots b_{d-2-s}}^{a_{1}\cdots a_{d-2-s}} \nonumber\\
&\qquad\qquad\qquad \times  \mathcal{R}%
_{\phantom{b_{2}b_{3}}a_{1}a_{2}}^{b_{1}b_{2}}\left[  \sigma\right]  \cdots\mathcal{R}%
_{\phantom{b_{2}b_{3}}a_{2m-1}a_{2m}}^{b_{2m-1}b_{2m}}\left[  \sigma\right]  \left(  k^{\left(
2\right)  }\right) _{a_{2m+1}}^{b_{2m+1}}\cdots\left(  k^{\left(  2\right)
}\right)  _{a_{d-2-s}}^{b_{d-2-s}} +\cdots \,,
\end{align}
using the FG-like expansions for the curvatures of $\partial\Sigma$ evaluated at the cutoff
radius $\rho=\epsilon$. Up to the next-to-leading order, we then have
\begin{align}
PI  &  =\left(  -1\right)  ^{\frac{d-3}{2}}\left(  d-3\right)  !\left(
\frac{1}{\ell_{\text{eff}}^{d-2}}- \epsilon\left(  \frac{\mathcal{R}^{\left(  0\right)  }}%
{2(d-4)\ell_{\text{eff}}^{d-4}}+\frac{1}{\ell_{\text{eff}}^{d-2}}\left(
tr\left[  \sigma^{\left(  2\right)  }\right]  +\frac{\ell_{\text{eff}}%
^{2}\kappa_{a}^{\left(  i\right)  a}\kappa_{b}^{\left(  i\right)  b}}%
{2(d-2)}\right)  \right)  \right)  + \cdots \,.
\end{align}
Noting that
\begin{equation}
\sqrt{\widetilde{\gamma}}=\frac{\sqrt{\sigma^{\left(  0\right)  }}}%
{\epsilon^{\frac{d-2}{2}}}\left(  1+\frac{\epsilon}{2}tr\left[  \sigma
^{\left(  2\right)  }\right]  +\mathcal{O}\left(  \epsilon^{2}\right)
\right)  ,
\end{equation}
we then obtain%
\begin{align}
I_{KT}^{\text{odd}}  &  =-\left(  \frac{d+1}{2}\right)  \left(  d-1\right)
c_{d}%
{\displaystyle\int\limits_{\partial\Sigma}}
d^{d-2}y\sqrt{\widetilde{\gamma}}PI\nonumber\\
&  =-%
{\displaystyle\int\limits_{\partial\Sigma}}
d^{d-2}y\frac{\sqrt{\sigma^{\left(  0\right)  }}}{\left(  d-2\right)
\epsilon^{\frac{d-2}{2}}}\sum\limits_{p=0}^{\frac{d-3}{2}}\frac{\left(
-1\right)  ^{p}\left(  p+1\right)  \left(  d-1\right)  !\alpha_{\left(
p+1\right)  }}{\left(  d-1-2p\right)  !\ell_{\text{eff}}^{2p-1}}%
\nonumber\\
&\qquad \qquad\qquad \times  \left(  1-\epsilon\left[  \frac{\ell_{\text{eff}}^{2}}{2(d-4)}%
\mathcal{R}^{\left(  0\right)  }+\frac{1}{2}tr\left(  \sigma^{\left(
2\right)  }\right)  +\frac{\ell_{\text{eff}}^{2}}{2(d-2)}\kappa_{a}^{\left(
i\right)  a}\kappa_{b}^{\left(  i\right)  b}\right]  \right)  + \cdots \,,
\end{align}
for odd $d$.

For even $d$, the expression for the Kounterterm is
\begin{align}
I_{KT}^{\text{even}}  &  =4GS_{KT}^{\text{even}}=\left\lfloor \frac{d+1}%
{2}\right\rfloor c_{d}%
{\displaystyle\int\limits_{\partial\Sigma}}
d^{d-2}yB_{d-2} \nonumber\\
&  =\left(  \frac{d}{2}\right)  c_{d}\left(  -\left(  d-1\right)  \right)
{\displaystyle\int\limits_{\partial\Sigma}}
d^{d-2}y\sqrt{\tilde{\gamma}}%
{\displaystyle\int\limits_{0}^{1}}
dt%
{\displaystyle\int\limits_{0}^{t}}
ds\delta_{b_{1}\cdots b_{d-3}}^{a_{1}\cdots a_{d-3}}k_{a_{1}}^{b_{1}}%
\nonumber\\
& \qquad \times  \left(  \frac{1}{2}\mathcal{R}_{\phantom{b_{2}b_{3}}a_{2}a_{3}}^{b_{2}b_{3}}-t^{2}k_{a_{2}%
}^{b_{2}}k_{a_{3}}^{b_{3}}+\frac{s^{2}}{\ell_{\text{eff}}^{2}}\delta_{a_{2}%
}^{b_{2}}\delta_{a_{3}}^{b_{3}}\right)  \cdots\left(  \frac{1}{2}%
\mathcal{R}_{\phantom{b_{2}b_{3}}a_{d-4}a_{d-3}}^{b_{d-4}b_{d-3}}-t^{2}k_{a_{d-4}}^{b_{d-4}%
}k_{a_{d-3}}^{b_{d-3}}+\frac{s^{2}}{\ell_{\text{eff}}^{2}}\delta_{a_{d-4}%
}^{b_{d-4}}\delta_{a_{d-3}}^{b_{d-3}}\right)  \,,
\end{align}
where for this case, the boundary term
$
{\displaystyle\int}
d^{d-2}yB_{d-2}$ was evaluated in ref.\cite{Anastasiou:2019ldc}, and the computational procedure is
analogous to the odd $d$ case. Hence  $I_{KT}^{\text{odd}} 
=I_{KT}^{\text{even}}$, and  the
Kounterterm $I_{KT}$ has the same form for both odd and even $d$
as given in eq.\eqref{I_Kt} of the main text.

\bibliographystyle{JHEP}
\bibliography{EELove}
\end{document}